\documentclass[twocolumn,english,superscriptaddress,floatfix]{revtex4}
\usepackage[T1]{fontenc}
\usepackage[latin9]{inputenc}
\usepackage{amsbsy}
\usepackage{graphicx}
\usepackage{esint}

\makeatletter
\@ifundefined{textcolor}{}
{%
 \definecolor{BLACK}{gray}{0}
 \definecolor{WHITE}{gray}{1}
 \definecolor{RED}{rgb}{1,0,0}
 \definecolor{GREEN}{rgb}{0,1,0}
 \definecolor{BLUE}{rgb}{0,0,1}
 \definecolor{CYAN}{cmyk}{1,0,0,0}
 \definecolor{MAGENTA}{cmyk}{0,1,0,0}
 \definecolor{YELLOW}{cmyk}{0,0,1,0}
}

\usepackage{babel}
\usepackage{babel}
\usepackage{babel}
\usepackage{babel}
\usepackage{babel}
\usepackage{babel}
\usepackage{babel}
\usepackage{babel}
\usepackage{babel}
\usepackage{babel}

\usepackage{babel}

\usepackage{babel}

\usepackage{babel}

\usepackage{babel}

\makeatother

\usepackage{babel}

\newcommand{\D}{\Delta}
\newcommand{\e}{\varepsilon}

\begin{document}

\title{How many quantum phase transitions exist inside the superconducting
dome of the iron pnictides?}

\author{Rafael M. Fernandes}

\affiliation{School of Physics and Astronomy, University of Minnesota, Minneapolis,
MN 55455, USA}

\author{Saurabh Maiti}

\affiliation{Department of Physics, University of Wisconsin-Madison, Madison,
Wisconsin 53706, USA}

\author{Peter Wölfle}

\affiliation{Institute for Condensed Matter Theory and Institute for Nanotechnology,
Karlsruhe Institute of Technology, 76021 Karlsruhe, Germany}

\author{Andrey V. Chubukov}

\affiliation{Department of Physics, University of Wisconsin-Madison, Madison,
Wisconsin 53706, USA}

\date{\today }
\begin{abstract}
Recent experiments on two iron-pnictide families suggest the existence
of a single quantum phase transition (QPT) inside the superconducting
dome despite the fact that two separate transition lines - magnetic
and nematic - cross the superconducting dome at $T_{c}$. Here we
argue that these two observations are actually consistent. We show,
using a microscopic model, that each order coexists with superconductivity
for a wide range of parameters, and both transition lines continue
into the superconducting dome below $T_{c}$. However, at some $T_{\mathrm{merge}}<T_{c}$,
the two transitions merge and continue down to $T=0$ as a single
simultaneous first-order nematic/magnetic transition. We show that
superconductivity has a profound effect on the character of this first-order
transition, rendering it weakly first-order and allowing strong fluctuations
to exist near the QPT. 
\end{abstract}


\maketitle
\emph{Introduction.\ } A common theme across different phase diagrams
of unconventional superconductors (SC) is the idea of one or more
continuous quantum phase transitions (QPT's) under the SC dome \cite{Sachdev10}.
Examples include heavy fermion materials \cite{Wolfle_RMP,Wolfle11},
cuprates \cite{Moon10}, and iron pnictides \cite{reviews}. Such
QPT is generally associated with a non-superconducting (SC) order
which penetrates into the SC dome \cite{Moon10,Chubukov03,Metlitski10,Zaanen11,Scalapino12,Efetov12,Chubukov13}.
Direct experimental access to this putative QPT requires killing the
SC order, which can be challenging in high-temperature superconductors
due to high value of their critical magnetic fields \cite{Sachdev10}.
An alternative is to search for the QPT directly inside the SC dome.
However, there is no guarantee that the non-SC continuous phase transition
persists down to $T=0$, as it may become first-order if the SC and
non-SC orders do not coexist microscopically \cite{Vorontsov09,FernandesPRB10,Vorontsov10,Fernandes_Schmalian}.

In the iron pnictides, measurements of the $T=0$ SC penetration depth
across the phase diagram of $\mathrm{BaFe_{2}\left(As_{1-x}P_{x}\right)_{2}}$
found a pronounced peak at $x\approx0.3$ inside the SC dome, consistent
with the existence of a single QPT \cite{Hashimoto12}. Because in
the phase diagram of this and other iron pnictides, e.g. $\mathrm{Ba\left(Fe_{1-x}Co_{x}\right)_{2}As_{2}}$,
a spin-density wave (SDW) transition line meets the SC dome near the
highest $T_{c}$ \cite{Kasahara10,FernandesPRB10}, it is natural
to identify the observed peak with a magnetic quantum critical point,
like in heavy fermions and other materials. However, in the iron pnictides
there is not only one, but two separate phase transition lines that
cross the SC dome \cite{Nandi10,Bohmer12}. Besides the SDW transition
at $T_{m}$, there is also a nematic/structural transition at $T_{n}>T_{m}$,
below which the tetragonal $C_{4}$ symmetry of the system is spontaneously
broken down to $C_{2}$ \cite{Fisher_RPP,Fernandes12}. This peculiar
feature raises the issue of how many QPTs - if any - exist inside
the SC dome.

\begin{figure}
\begin{centering}
\includegraphics[width=0.8\columnwidth]{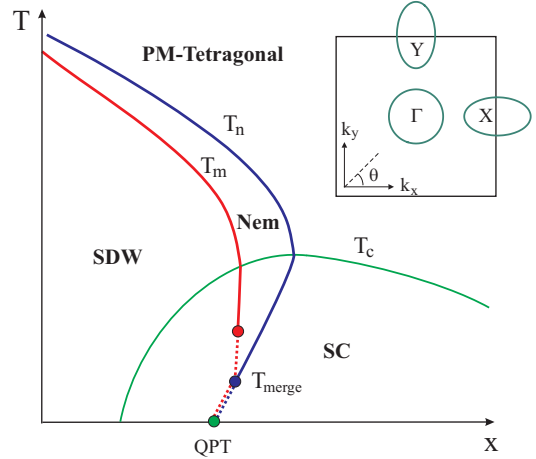}
\par\end{centering}

\caption{Schematic phase diagram summarizing our main results. The spin-density
wave and nematic transition lines, $T_{m}$ and $T_{n}$, separately
cross the superconducting transition line, $T_{c}$, and coexist with
superconductivity immediately below $T_{c}$. As temperature is lowered,
one of the transitions becomes first-order (dashed line). At a smaller
$T=T_{\mathrm{merge}}$ the two transition lines merge onto a single
simultaneous weakly first-order transition, which persists down to
$T=0$ and gives rise to a single QPT inside the dome. The back-bending
of the lines below $T_{c}$ may or may not take place (see Refs. \cite{Fernandes_Schmalian,Vorontsov10,Moon10}).
Inset: schematic Fermi surface. \label{fig_general_phase_diagram}}
\end{figure}

In this paper we address this issue by using a microscopic electronic
model which describes simultaneously the magnetic, nematic, and superconducting
phases. We show that SDW and nematic orders coexist with superconductivity
such that both the $T_{m}$ and $T_{n}$ lines penetrate separately
into the SC dome. However, at $T=0$, deep inside the SC dome, the
nematic and SDW transition lines merge on a\emph{ single weakly first-order}
QPT (Fig. \ref{fig_general_phase_diagram}). The weak character of
this transition is a direct consequence of the coexistence with SC,
and implies the persistence of quantum critical fluctuations for wide
temperature and doping ranges, which should affect the macroscopic
properties of these materials. Our results also reconcile the existence
of two split phase transitions above $T_{c}$ with the penetration
depth measurements of \cite{Hashimoto12}, which point to a single
phase transition at $T=0$ inside the SC dome.

\emph{Microscopic model.\ }We consider a minimal model consisting
of one circular hole pocket at the center of the 1-Fe Brillouin zone
and two elliptical electron pockets centered at momenta $\mathbf{Q}_{X}=\left(\pi,0\right)$
and $\mathbf{Q}_{Y}=\left(0,\pi\right)$. The band dispersions are
parameterized in terms of the momentum $k$ and angle $\theta$ as
$\varepsilon_{\Gamma,\mathbf{k}}=-\varepsilon_{k}=\varepsilon_{0}-\frac{k^{2}}{2m}$,
$\varepsilon_{X,\mathbf{k}}=\varepsilon_{\mathbf{k-Q_{X}}}+2\delta_{0}+2\delta_{2}\cos2\theta$,
and $\varepsilon_{Y,\mathbf{k}}=\varepsilon_{\mathbf{k-Q_{Y}}}+2\delta_{0}-2\delta_{2}\cos2\theta$,
where $\delta_{0}$ is proportional to doping and $\delta_{2}$ originates
from the ellipticity of the electron pockets \cite{Eremin10,Fernandes12}
(see inset in Fig. 1). The Hamiltonian of the model is $H=H_{2}+H_{4}$.
The free-fermion part is $H_{2}=\sum_{a,\mathbf{k}}\left(\varepsilon_{a,\mathbf{k}}-\mu\right)c_{a,\mathbf{k}\sigma}^{\dagger}c_{a,\mathbf{k}\sigma}^{\phantom{\dagger}}$,
where $\sigma$ is a spin index, $a$ is a band index, and $\mu$
is the chemical potential. The interaction term $H_{4}$ contains
eight different 4-fermion interactions \cite{Maiti10}. 

We follow earlier works and assume that SDW magnetism with momentum
$\mathbf{Q}_{X}$ and/or $\mathbf{Q}_{Y}$ (order parameters ${\bf M}_{j}=\sum_{\mathbf{k}}c_{\Gamma,\mathbf{k}\alpha}^{\dagger}\boldsymbol{\sigma}_{\alpha\beta}c_{j\mathbf{,k+Q_{j}}\beta}^{\phantom{\dagger}}$,
$j=X,Y$) and $s^{+-}$ superconductivity (order parameters $\Delta_{i}=\sum_{\mathbf{k}}c_{i,\mathbf{k}\uparrow}^{\dagger}c_{i,-\mathbf{k}\downarrow}^{\dagger}$,
$i=X,Y,\Gamma$) are the primary instabilities, while nematicity is
caused by magnetic fluctuations \cite{Kivelson,Sachdev,FernandesPRL10,Fernandes12}.
Decoupling the interaction terms in $H_{4}$ and integrating over
the fermions, we obtain the effective action $S_{\mathrm{eff}}\left[\Delta_{i},M_{j}\right]$
and expand in $\Delta$ and $M_{i}$: 
\begin{eqnarray}
 &  & S_{\mathrm{eff}}=a_{m}\left(M_{X}^{2}+M_{Y}^{2}\right)+a_{s}\Delta^{2}+\frac{u_{m}}{2}\left(M_{X}^{2}+M_{Y}^{2}\right)^{2}\nonumber \\
 &  & -\frac{g_{m}}{2}\left(M_{X}^{2}-M_{Y}^{2}\right)^{2}+\frac{u_{s}}{2}\Delta^{4}+\lambda\Delta^{2}\left(M_{X}^{2}+M_{Y}^{2}\right)\label{action}
\end{eqnarray}
with coefficients depending on the interactions and the band parameters
$\left(\delta_{0},\delta_{2}\right)$ (see Supplementary Material
(SM) for details). The coefficients $a_{m}$ and $a_{s}$ vanish at
the mean-field SDW and SC transition temperatures $T_{m,0}$ and $T_{c,0}$,
while the coefficients $u_{m}>g_{m},u_{s}$, and $\lambda$ are all
positive at not very low $T$. Here we considered equal inter-band
pairing interactions, implying $\Delta_{\Gamma}=-\sqrt{2}\Delta_{X,Y}=\Delta$.

\begin{figure}
\includegraphics[width=0.5\columnwidth]{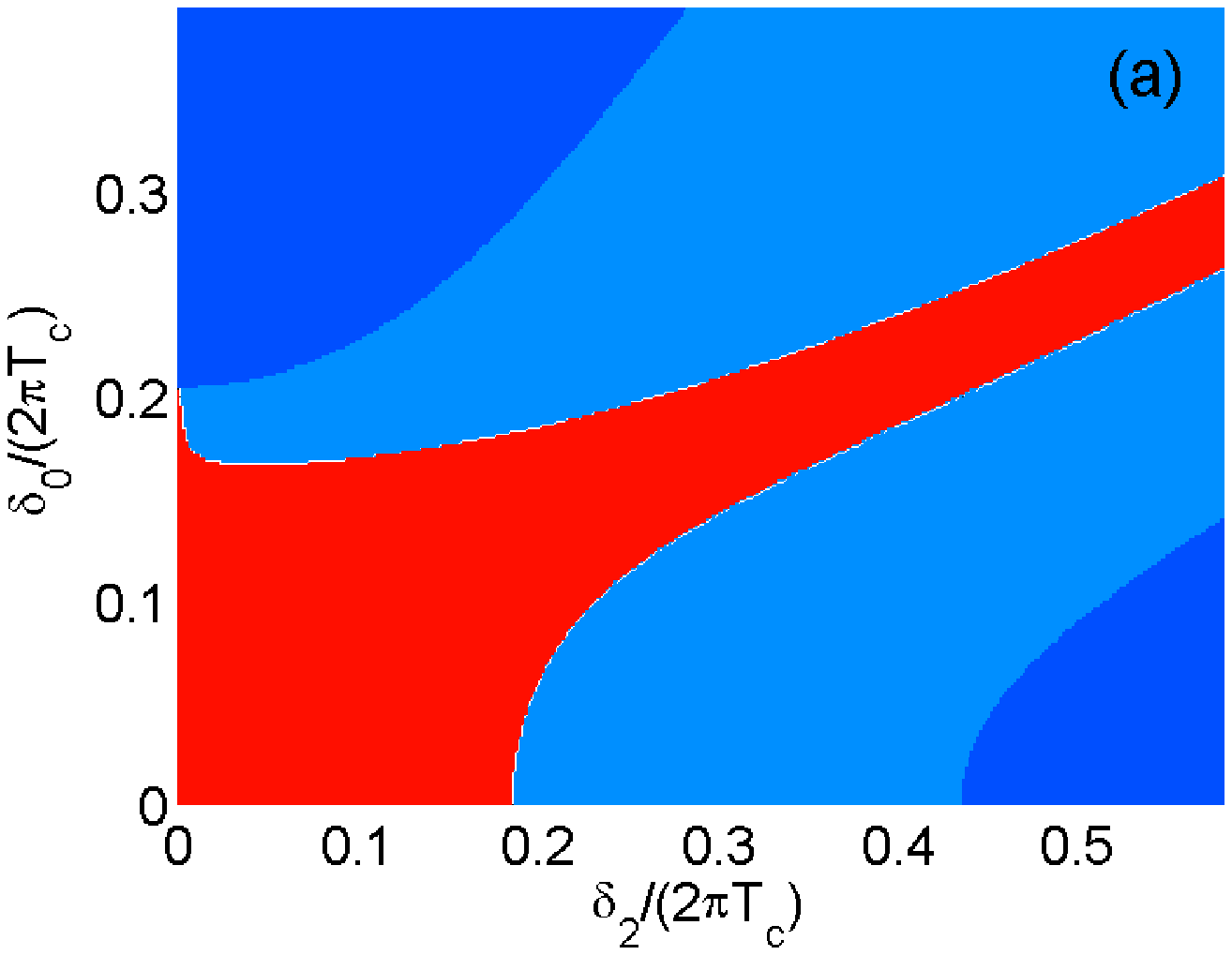}\hfill{}\includegraphics[width=0.5\columnwidth]{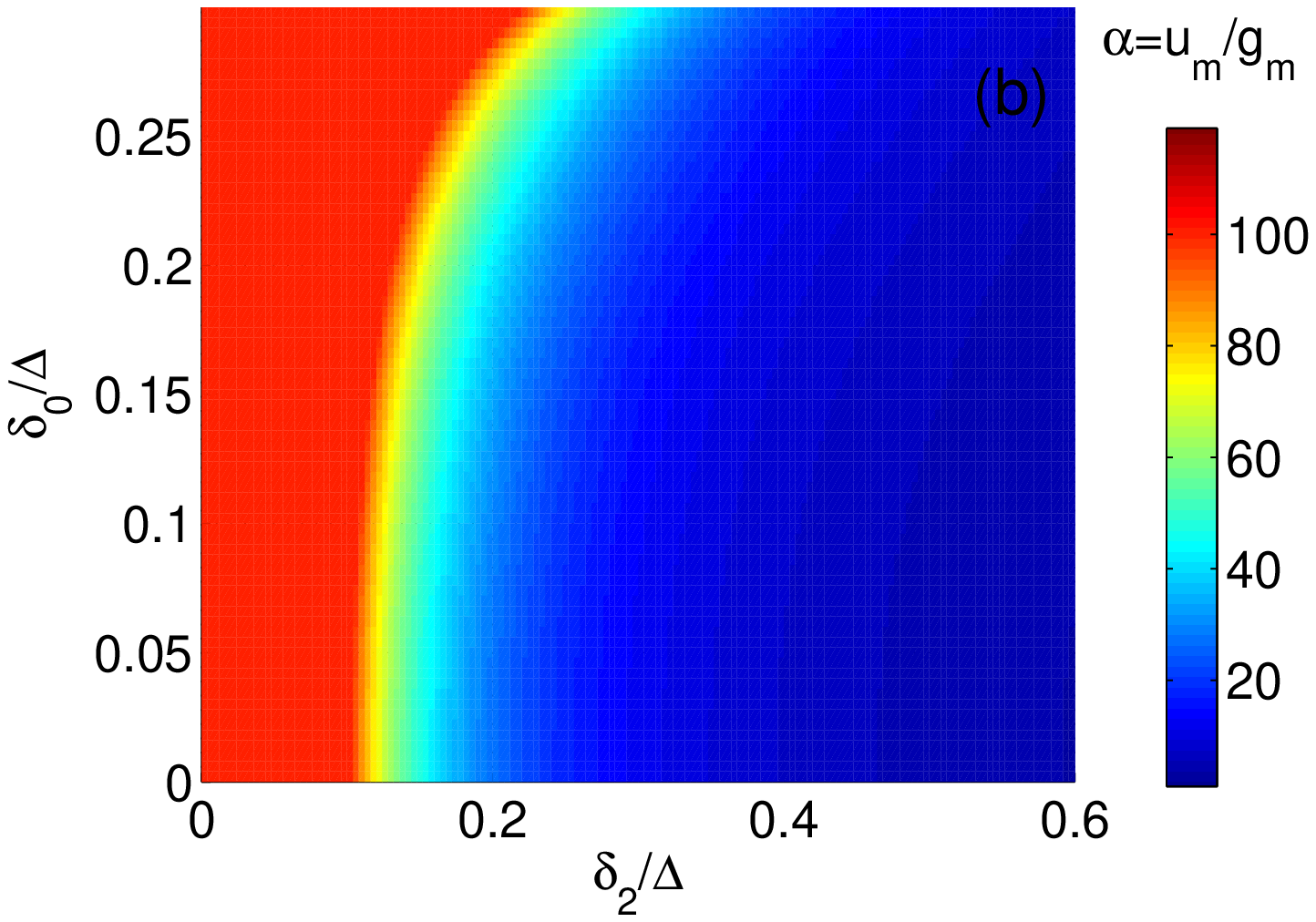}
\caption{ (a) The regions of coexistence between mean-field SDW and superconductivity
(light-blue region) and nematicity and superconductivity (red region)
in the $\left(\delta_{0},\delta_{2}\right)$ parameter space, in $d=2$.
In the red region both SDW and nematic order coexist with SC order.
(b) Color plot of $\alpha=u_{m}/g_{m}$ inside the SC state for different
$\delta_{0}/\Delta$ and $\delta_{2}/\Delta$ ($\Delta$ is the SC
gap). \label{fig_coexistence}}
\end{figure}

\emph{Mean-field analysis.\ }We first analyze this action at the
mean-field level, when the $M$ and $\Delta$ fields do not fluctuate.
In this case, although there is no preemptive nematic order, the tetragonal
symmetry is broken below $T_{m,0}$, because the minimum of Eq. (\ref{action})
is a stripe SDW phase with either $M_{X}=0$, $M_{Y}\neq0$ or $M_{X}\neq0,M_{Y}=0$.
The competing SC and SDW orders coexist microscopically as long as
the quartic coefficients satisfy the condition $\lambda<\sqrt{u_{s}\left(u_{m}-g_{m}\right)}$
(Refs.\cite{Vorontsov09,FernandesPRB10,Vorontsov10,Fernandes_Schmalian}).
This happens in the light-blue region of the $\left(\delta_{0},\delta_{2}\right)$
space of Fig. \ref{fig_coexistence}. \cite{comm_a}For parameters
in this range, the continuous SDW transition line penetrates into
the SC state, albeit with a different slope~\cite{Moon10}. It survives
down to $T=0$ if $u_{m}>0$, which is the case when the SC gap is
not too small (see SM). In this case, the mean-field SDW transition
remains second-order down to $T=0$ and ends up at a magnetic quantum
critical point under the SC dome.

\emph{Preemptive nematic order.\ }To include fluctuations of the
SDW fields $M_{X}$ and $M_{Y}$ we replace $a_{m}$ in Eq. (\ref{action})
by the SDW susceptibility $\chi_{0}^{-1}\left(\mathbf{Q}_{i}+\mathbf{q},\omega_{n}\right)=a_{m}+q^{2}+f\left(\omega_{n}\right)$,
where $\omega_{n}=2\pi nT$ is the Matsubara frequency and $f\left(\omega_{n}\right)$
is proportional to $\left|\omega_{n}\right|$ in the normal state
and $\omega_{n}^{2}$ deep inside the SC dome. We then introduce two
Hubbard-Stratonovich fields $\psi=u_{m}\left\langle M_{X}^{2}+M_{Y}^{2}\right\rangle $
and $\varphi=g_{m}\left\langle M_{X}^{2}-M_{Y}^{2}\right\rangle $,
integrate the partition function over $M_{i}(\mathbf{q},\omega_{n})$,
and obtain the effective action $S_{\mathrm{eff}}\left[\Delta,\psi,\varphi\right]$.
Fluctuations of $M_{i}$ and of $\psi$ and $\varphi$ are conjugated
- if $M_{i}$ fluctuates strongly, as we now assume, fluctuations
of $\psi$ and $\varphi$ are weak, and the effective action $S_{\mathrm{eff}}\left[\Delta,\psi,\varphi\right]$
can in turn be analyzed in the saddle-point approximation (see SM).
The field $\left\langle \psi\right\rangle $ is always non-zero and
shifts the ``pure'' SDW transition temperature from $T_{m,0}$ down
to $\tilde{T}_{m,0}$. Our analysis, for which we used the expansion
to order $M^{4}$ in Eq. (\ref{action}) is valid when $(T_{n}-\tilde{T}_{m,0})/T_{n}\leq1$.
A non-zero $\left\langle \varphi\right\rangle $ appears only below
a certain $T_{n}$ and breaks the tetragonal $C_{4}$ symmetry down
to $C_{2}$, inducing an orthorhombic distortion and orbital order~\cite{Fernandes12}.
If $\left\langle \varphi\right\rangle $ becomes non-zero at $T_{n}>\tilde{T}_{m,0}$,
there exists a temperature range in which the system displays nematic
order $\left\langle \varphi\right\rangle \neq0$ but no long-range
magnetic order $\left\langle {\bf M}_{i}\right\rangle =0$. In the
normal state, the effective action $S_{\mathrm{eff}}\left[0,\psi,\varphi\right]$
is 
\begin{equation}
S_{\mathrm{eff}}\left[0,\psi,\varphi\right]=\frac{\varphi^{2}}{2g_{m}}-\frac{\psi^{2}}{2u_{m}}+\frac{3}{2}\int_{q}\log{\left[\left(\chi_{0}^{-1}+\psi\right)^{2}-\varphi^{2}\right]}\label{sa_1}
\end{equation}
 where $\int_{q}=T\sum_{\omega_{n}}\int d^{d}q/(2\pi)^{d}$. This
action has been analyzed before in several contexts\cite{Fernandes12,Gorkov09,Qi09,Cano10,batista,Applegate11,lorenzana,millis10}.
For quasi-2D layered systems, the behavior depends on the ratio $\alpha=u_{m}/g_{m}\geq1$.
For $\alpha$ relevant to near-optimally doped $\mathrm{BaFe_{2}\left(As_{1-x}P_{x}\right)_{2}}$
and $\mathrm{Ba\left(Fe_{1-x}Co_{x}\right)_{2}As_{2}}$, the nematic
transition is second order and occurs at $T_{n}>\tilde{T}_{m,0}$,
i.e. before the ``pure'' SDW transition. A non-zero $\left\langle \varphi\right\rangle $
shifts the SDW transition upwards from $\tilde{T}_{m,0}$ to $T_{m}$,
but still $T_{m}<T_{n}$. In this situation, there are two split second-order
transition lines, $T_{n}$ and $T_{m}$, which separately cross the
$T_{c}$ line. Our goal now is to find the fate of these transitions
inside the SC dome.

\emph{Coexistence of nematicity and SC.\ }We first consider the vicinity
of the point where the nematic transition line $T_{n}$ hits $T_{c}$.
For simplicity, we set $d=2$ to study the coexistence between nematicity
and SC. This procedure is safe for the nematic order, as it only breaks
a discrete symmetry. Inter-layer coupling will only account for small
corrections to $T_{n}$, but it is crucial for the existence of an
SDW transition line at $\tilde{T}_{m,0}>0$. We assume that $T_{c}$
is large enough and neglect the dynamic part of $\chi_{0}$. We obtain
$\psi$ from the saddle-point equation $\partial S_{\mathrm{eff}}/\partial\psi=0$,
substitute the result back into the effective action and obtain $S_{\mathrm{eff}}\left[\Delta,\varphi\right]$:

\begin{equation}
S_{\mathrm{eff}}\left[\Delta,\varphi\right]=a_{n}\varphi^{2}+\frac{u_{n}}{2}\varphi^{4}+\tilde{a}_{s}\Delta^{2}+\frac{\tilde{u}_{s}}{2}\Delta^{4}+\tilde{\lambda}\varphi^{2}\Delta^{2}\label{S_nem}
\end{equation}
 where : 
\begin{equation}
\tilde{\lambda}=\frac{\lambda}{2(u_{m}+g_{m})},~\tilde{u}_{s}=\frac{u_{s}-\frac{\lambda^{2}}{u_{m}+g_{m}}}{2g},~u_{n}=\frac{1}{6}\frac{u_{m}-2g_{m}}{u_{m}+g_{m}}
\end{equation}

Notice that all coefficients originate from the SDW/SC action (\ref{action}),
i.e. the coupling between the nematic and SC order parameters is a
consequence of the coupling between the SDW and SC fields ($\tilde{\lambda}\propto\lambda$)~\cite{FernandesPRL10}.
In the absence of SC, the nematic transition is second-order when
$u_{n}>0$, i.e. $\alpha=u_{m}/g_{m}>2$, which we assume to hold.

It follows from Eq. (\ref{S_nem}) that nematic and SC orders coexist
when $\tilde{\lambda}<\sqrt{\tilde{u}_{s}u_{n}}$, which in terms
of the original Ginzburg-Landau coefficients gives $\lambda<\sqrt{u_{s}\left(u_{m}-2g_{m}\right)}$.
Although this is a more restrictive condition than $\lambda<\sqrt{u_{s}\left(u_{m}-g_{m}\right)}$
for the coexistence between mean-field SDW and SC, it is still satisfied
in a rather wide range of parameters $\left(\delta_{0},\delta_{2}\right)$,
including the region of small $\delta_{0}$ and $\delta_{2}$ (the
red region in Fig. \ref{fig_coexistence}a). In this parameter range,
the second-order $T_{n}$ line continues below $T_{c}$, albeit with
a different slope. Because the condition for SDW-SC coexistence is
the same both in mean-field and in the presence of Gaussian fluctuations~\cite{FernandesPRB10},
in the same red region of Fig. \ref{fig_coexistence}a, the SDW $T_{m}$
line also continues as a second-order transition line into the SC
dome.

\emph{Nematic and SDW transitions at $T=0$.\ } At $T=0$, the dynamics
of $\chi_{0}\left(\mathbf{q},\omega_{n}\right)$ cannot be neglected.
Deep in the SC state, the spin dynamics is propagating $\chi_{0}^{-1}\left(\mathbf{Q}_{i}+\mathbf{q},\omega_{n}\right)=a_{m}+q^{2}+\omega_{n}^{2}$,
i.e. the quantum system behaves like the classical system in an effective
dimension $d_{\mathrm{eff}}=d+1=3$.

The effective action in terms of $\psi$ and $\varphi$ has the same
form as in the absence of SC, Eq. (\ref{sa_1}), but with renormalized
coefficients and in $d_{\mathrm{eff}}=3$. Anticipating that nematic
transition may trigger an instantaneous magnetic transition, we introduce
an SDW order parameter $m$ (the average value of either $M_{X}$
or $M_{Y}$, depending on the sign of $\varphi$) and write $S_{\mathrm{eff}}$
in terms of $\psi,\varphi$, and $m$ (Ref. \cite{Fernandes12}).
We again use $\partial S_{\mathrm{eff}}/\partial\psi=0$ to eliminate
$\psi$ and obtain the action in terms of $\varphi$ and $m$: 
\begin{eqnarray}
S_{\mathrm{eff}}\left[\varphi,m\right] & = & \frac{\varphi^{2}}{2g_{m}}-\frac{r\left(r-2\bar{a}_{m}\right)}{2u_{m}}+\frac{3}{2\pi}m^{2}\left(r-\left|\varphi\right|\right)\nonumber \\
 &  & -\frac{\left(r+\varphi\right){}^{3/2}+\left(r-\varphi\right){}^{3/2}}{4\pi}\label{sa2}
\end{eqnarray}
 where $r={\bar{a}}_{m}-(3u_{m}/8\pi)\left(\sqrt{r+\varphi}+\sqrt{r-\varphi}\right)+\left(3u_{m}/2\pi\right)m^{2}$
is a function of $\varphi$ and $m$, and ${\bar{a}}_{m}=a_{m}+(3\Lambda u_{m})/(2\pi^{2})$
is the renormalized distance to the $T=0$ SDW transition in the absence
of nematicity. The magnetic order parameter $m$ satisfies the equation
of state $m\left(r-\left|\varphi\right|\right)=0$. It vanishes if
the nematic order parameter either emerges continuously or jumps to
a value $\left|\varphi\right|<r$, but can become non-zero if $\varphi$
jumps at the nematic transition to $\left|\varphi\right|=r$. That
$m$ can become non-zero right at the nematic transition can also
be understood by looking at the SDW susceptibility $\chi\left(\mathbf{Q}_{i}\right)\propto1/r$.
For $\varphi=0$, the SDW susceptibility diverges when $r=0$, which
happens at ${\bar{a}}_{m}=0$. If the nematic transition occurs at
${\bar{a}}_{m}>0$, preempting the magnetic transition, the static
SDW susceptibility splits into $\chi\left(\mathbf{Q}_{i}\right)\propto1/(r\pm\varphi)$.
If $\varphi$ jumps to $\left|\varphi\right|=r$ at the nematic transition,
one of the $\chi\left(\mathbf{Q}_{i}\right)$ diverges, and $m$ may
also jump.

\begin{figure}
\begin{centering}
\includegraphics[width=0.7\columnwidth]{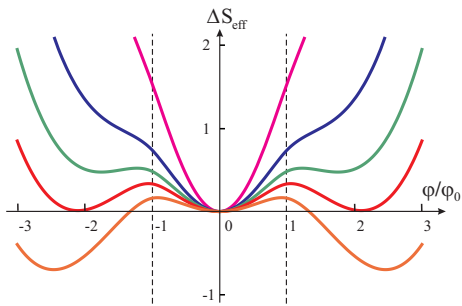} 
\par\end{centering}

\caption{ The effective action $S_{\mathrm{eff}}\left[\varphi\right]$ at $T=0$
and $d_{\mathrm{eff}}=3$, as a function of the nematic order parameter
$\varphi$ for various $\bar{a}_{m}$, which measure the distance
to the pure SDW $T=0$ transition. From top to bottom, $\bar{a}_{m}\left(\frac{4\pi^{2}}{9u_{m}^{2}}\right)=0.0345$,
$0.0320$, $0.0310$, $0.03034$, and $0.0295$. The dashed line $\varphi=\varphi_{0}$
separates the region where nematic order does not induce magnetic
order ($\left|\varphi\right|<\varphi_{0}$) from the regions where
magnetic order is simultaneously induced ($\left|\varphi\right|>\varphi_{0}$).
We set $u_{m}/g_{m}=5$. \label{fig_free_energy}}
\end{figure}

We analyzed $S_{\mathrm{eff}}\left[\varphi,m\right]$ by reducing
${\bar{a}}_{m}$ from some initially large positive value in the paramagnetic
phase down to $\bar{a}_{m}=0$ (at the pure $T=0$ SDW transition).
This is valid for systems where the pure SDW transition is continuously
suppressed to zero. In the range, $\left|\varphi\right|\leq\varphi_{0}$,
where $\varphi_{0}=\frac{1}{32}\left(\sqrt{\frac{9u_{m}^{2}}{4\pi^{2}}+32\bar{a}_{m}}-\frac{3u_{m}}{2\pi}\right)^{2}$,
we have $\left|\varphi\right|\leq r$ and hence $m=0$. For $\left|\varphi\right|>\varphi_{0}$,
we have $r=\left|\varphi\right|$ and $m(\varphi,{\bar{a}}_{m})\neq0$
determined from the equation on $r$. Our results are shown in Fig.
\ref{fig_free_energy} where we plotted $S_{\mathrm{eff}}\left[\varphi\right]$
in both regions at various ${\bar{a}}_{m}$. For large ${\bar{a}}_{m}$,
$S_{\mathrm{eff}}(\varphi)$ has a minimum at $\varphi=0$ and monotonically
increases with $|\varphi|$. When $\bar{a}_{m}$ becomes smaller than
$\bar{a}_{m,c1}=(3u/4\pi)^{2}/\left(2(\alpha-1)\right)$, $S_{\mathrm{eff}}(\varphi)$
develops inflection points at $\left|\varphi\right|>\varphi_{0}$.
Upon decreasing $\bar{a}_{m}$ further, these inflection points split
in two pairs of local maximum and minimum $\varphi=\pm\varphi_{\mathrm{max}}$
and $\varphi=\pm\varphi_{\mathrm{min}}$. At some $0<\bar{a}_{m,cr}\leq\bar{a}_{m,c1}$,
$S_{\mathrm{eff}}\left[\varphi=\pm\varphi_{\min}\right]$ eventually
becomes lower than $S_{\mathrm{eff}}\left[\varphi=0\right]$, i.e.
the system undergoes a first-order nematic transition in which the
nematic order parameter jumps from $\varphi=0$ to $\varphi=\pm\varphi_{\mathrm{min}}$.
Because $\varphi_{\mathrm{min}}>\varphi_{0}$, the jump in the nematic
order parameter is strong enough to induce a simultaneous first-order
magnetic transition. Since at the transition both $\varphi$ and $m$
jump simultaneously to finite values, there is only one first-order
QPT under the SC dome (see Fig. \ref{fig_general_phase_diagram}).

We verified that at ${\bar{a}}_{m,cr}$ the coefficient of the $\varphi^{2}$
term in $S_{\mathrm{eff}}[\varphi]$ remains positive for all $\alpha\equiv u_{m}/g_{m}$,
i.e. the first-order nematic transition preempts not only the SDW
transition but also the potential second-order nematic transition.
This result is at variance with earlier works (Ref. \cite{Qi09})
which suggested separate second-order transitions at $T=0$ in $d_{\mathrm{eff}}=3$.

An important issue is the strength of this first-order transition.
For $d_{\mathrm{eff}}=3$, $\varphi_{0}$ and the jump in the nematic
order parameter $\delta\phi$ decreases when $\alpha$ increases and
scale as $1/\alpha^{2}$ at large $\alpha$. Similarly, $\bar{a}_{m,c1}$,
below which the first-order transition preempts the pure SDW QCP,
scales as $1/\alpha$. These scalings are a general consequence of
the fact that $d_{\mathrm{eff}}=3$ is the borderline between the
regimes of simultaneous and split nematic/SDW transitions, since for
$d_{\mathrm{eff}}=3-\epsilon$ the two transition become split and
second-order for $\alpha>\frac{3}{2\epsilon}$. Interestingly, we
found that, in a wide region of $\left(\delta_{0},\delta_{2}\right)$,
$\alpha$ becomes large in the SC state (see Fig. \ref{fig_coexistence}b),
i.e. the presence of superconductivity makes the first-order transition
weaker. Note in this regard that the effective dimension $d_{\mathrm{eff}}=3$
is also a direct consequence of the presence of SC, which changes
the spin dynamics to propagating. Without SC, the spin dynamics would
be diffusive with $d_{\mathrm{eff}}=2+z=4$, and the first-order transition
would be much stronger \cite{comm_aa}.

The weak character of the first-order QPT inside the SC dome is also
manifested in the temperature range $0<T<T_{c}$. By combining the
present results at $T=0$ and near $T_{c}$ with the earlier analysis
of the classical phase diagram in quasi-2D systems \cite{Fernandes12},
we find that the nematic and magnetic transition lines merge at some
non-zero temperature $T_{\mathrm{merge}}<T_{c}$, below which the
two orders develop simultaneously via a first-order transition (see
Fig. \ref{fig_general_phase_diagram}). The details of the phase diagram
near $T_{\mathrm{merge}}$ depend on the strength of the inter-layer
coupling, with either the nematic or the magnetic transition line
becoming second-order immediately above $T_{\mathrm{merge}}$. Most
importantly, $T_{\mathrm{merge}}$ also scales as $1/\alpha^{2}$
and is small at large $\alpha=u_{m}/g_{m}$. As a result, the system
behaves almost like the nematic and SDW second-order transition lines
would merge right at $T=0$. In this special case, stripe and non-stripe
magnetic states are degenerate, and the SDW order parameter manifold
is enhanced to $O(6)$, i.e. only the modulus of the 6-component vector
$({\bf M}_{X},{\bf M}_{Y})$ is fixed. This gives rise to enhanced
quantum fluctuations near the QPT. 

\emph{Comparison with experiments\ } Our results can be directly
applied to iron pnictides, particularly to $\mathrm{BaFe_{2}\left(As_{1-x}P_{x}\right)_{2}}$
and $\mathrm{Ba\left(Fe_{1-x}Co_{x}\right)_{2}As_{2}}$, whose SC
domes are crossed by two split second-order magnetic and nematic transition
lines. Microscopic coexistence between SDW and superconductivity has
been established in both cases by NMR \cite{NMR_coexistence_Co1,NMR_coexistence_Co2,NMR_coexistence_P},
and a suppression of the orthorhombic order parameter (proportional
to $\varphi$ in our model) has been found inside the SC dome \cite{Nandi10,Bohmer12}.
Our calculations predict a single simultaneous weak first-order nematic/SDW
QPT at $T=0$, which can be verified by measurements of the $T=0$
SC penetration depth. In $\mathrm{BaFe_{2}\left(As_{1-x}P_{x}\right)_{2}}$,
a single peak in the penetration depth has been observed near optimal
doping \cite{Hashimoto12}. Experiments cannot resolve whether it
implies a second-order or weakly first-order transition, but the fact
is that there is a single transition at $T=0$, despite two split
transitions crossing into the SC dome, in agreement with our theory.
The peak in the penetration depth (but not a divergence) is expected
due to $O(3)$ SDW fluctuations~\cite{Levchenko13,Sachdev13}. When
$T_{\mathrm{merge}}$ is small, as it is for large $\alpha$ (see
Fig. \ref{fig_coexistence}b), the emerging $O(6)$ symmetry further
enhances the strength of the peak. In $\mathrm{Ba\left(Fe_{1-x}Co_{x}\right)_{2}As_{2}}$,
penetration depth measurements have so far not identified a peak inside
the SC dome, yet the penetration depth was found to increase below
a certain doping \cite{Gordon10}. This increase is expected in the
SDW+SC phase due to the competition between SC and SDW orders \cite{Fernandes_Schmalian2,Vavilov_penetration_depth},
and in this regard the experimental result is again consistent with
the existence of a single transition point at $T=0$.

To summarize, in this paper we considered the behavior of the SDW
and nematic transition lines inside the SC dome. We argued that both
orders coexist with SC, and the two transition lines separately penetrate
into the SC dome as continuous second-order transitions. However,
as temperature is lowered, they merge at some small but finite $T_{\mathrm{merge}}$,
giving rise to a single nematic/SDW weakly first-order QPT. The weak
character of the transition is a direct consequence of the coexistence
with the SC order, which makes the spin dynamics propagating and enhances
the ratio of the quartic couplings, pushing the system to the borderline
between the first-order and second-order regimes.

We thank I. Eremin, M. Khodas, Y.~Matsuda, R. Prozorov, S.~Sachdev,
J. Schmalian, T.~Shibauchi, and O. Starykh for fruitful discussions.
A.V.C. and S.M. are supported by the DOE grant DE-FG02-ER46900. PW
is grateful for the hospitality extended to him as a visiting professor
at the University of Wisconsin, Madison, and acknowledges support
through an ICAM Senior Scientist Fellowship. SM acknowledges support
from ICAM-DMR-084415.

\newpage

\begin{widetext}

{\bf\Large \center Supplementary material for ``How many quantum phase transitions exist
inside the superconducting dome of the iron pnictides?"} \\

\setcounter{equation}{0}
\renewcommand{\theequation}{S\arabic{equation}}

\setcounter{figure}{0}
\renewcommand{\thefigure}{S\arabic{figure}}

\section{Derivation of the effective actions}

\subsection{Spin-density wave and superconductivity \label{sec:Model}}

We consider a model with one hole pocket at the $\Gamma-$point and
two symmetry-related elliptical electron pockets at the $X$ and $Y$
points of the unfolded Brilliouin Zone (BZ). Of the eight electronic
interactions present in this model \cite{S_Maiti10}, two contribute
to the superconducting (SC) and spin-density wave (SDW) channels:
the electron-hole density-density interactions ($U_{1}$) and electron-hole
pair hopping interactions ($U_{3}$). The Hamiltonian is given by
$H=H_{2}+H_{4}$, with:

\begin{eqnarray}
H_{2} & = & \sum_{\mathbf{k},i\in(X,Y,\Gamma)}\e_{\mathbf{k},i}c_{\mathbf{k}\sigma,i}^{\dag}c_{\mathbf{k}\sigma,i}\nonumber \\
H_{4} & = & \sum_{\mathbf{k},i\in(X,Y)}U_{1}c_{\mathbf{k}\alpha,\Gamma}^{\dag}c_{\mathbf{k}\gamma,i}^{\dag}c_{\mathbf{k}\delta,i}c_{\mathbf{k}\beta,\Gamma}\delta_{\alpha\beta}\delta_{\gamma\delta}\nonumber \\
 & + & \sum_{k,i\in(X,Y)}\frac{U_{3}}{2}\left(c_{\mathbf{k}\alpha,\Gamma}^{\dag}c_{\mathbf{k}\gamma,\Gamma}^{\dag}c_{\mathbf{k}\delta,i}c_{\mathbf{k}\beta,i}\,+\mathrm{h.c.}\right)\delta_{\alpha\beta}\delta_{\gamma\delta}
\end{eqnarray}
 where summation over spin indices is implied. The dispersions $\varepsilon_{\mathbf{k},i}$
are given in the main text as function of $\delta_{0}$ (which is
proportional to the chemical potential) and $\delta_{2}$ (which is
proportional to the ellipticity of the electron pockets). In $H_{4}$
we retain terms only in the spin and the pairing sector and define
the staggered spin operators $\mathbf{S}_{i}=\frac{1}{\sqrt{2}}\sum_{\mathbf{k}}c_{\mathbf{k}\alpha,\Gamma}^{\dag}\boldsymbol{\sigma}_{\alpha\beta}c_{\mathbf{k}\beta,i}$
and the pairing operators $b_{i}=\sum_{\mathbf{k}}c_{\mathbf{k}\uparrow,i}c_{-\mathbf{k}\downarrow,i}$.
We can then rewrite $H_{4}$ as:

\begin{equation}
H_{4}=-(U_{1}+U_{3})\sum_{i\in(X,Y)}\mathbf{S}_{i}\cdot\mathbf{S}_{i}+2U_{3}\sum_{i\in(X,Y)}\left(b_{\Gamma}b_{i}+\mathrm{h.c.}\right)\label{eq:H_int_final}
\end{equation}

After introducing the Hubbard-Stratonovich fields $\mathbf{M}_{(X,Y)}$
for $\mathbf{S}_{(X,y)}$, $\Delta_{h}$ for $b_{\Gamma}$, and $\Delta_{e}$
for $b_{(X,Y)}$, we obtain the action $S$ as function of the fermionic
fields as well as the SDW and SC fluctuating fields (assumed to be
homogeneous):

\begin{equation}
S\left[\Psi,\mathbf{M}_{i},\Delta_{i}\right]=\frac{2}{(U_{1}+U_{3})}\left(M_{X}^{2}+M_{Y}^{2}\right)-\frac{4}{U_{3}}\Delta_{h}\Delta_{e}-\int_{k}\hat{\Psi}_{k}^{\dagger}\left(i\omega_{n}-\hat{\mathcal{H}}_{\mathbf{k}}\right)\hat{\Psi}_{k}\label{action_formal}
\end{equation}

Here, $k=\left(\omega_{n},\mathbf{k}\right)$, with $\omega_{n}=\left(2n+1\right)\pi T$
denoting the fermionic Matsubara frequency, and $\int_{k}=T\sum_{\omega_{n}}\int\frac{d^{d}k}{(2\pi)^{d}}$.
The 12-dimensional Nambu operator is given by $\hat{\Psi}_{k}^{\dagger}=\left(\begin{array}{ccc}
\psi_{k,\Gamma}^{\dagger} & \psi_{k,X}^{\dagger} & \psi_{k,Y}^{\dagger}\end{array}\right)$, with:

\begin{equation}
\psi_{k,i}^{\dagger}=\left(\begin{array}{cccc}
c_{\mathbf{k}\uparrow,i}^{\dagger} & c_{\mathbf{k}\downarrow,i}^{\dagger} & c_{-\mathbf{k}\uparrow,i} & c_{-\mathbf{k}\downarrow,i}\end{array}\right)\label{eq:psi}
\end{equation}
 and:

\begin{equation}
\mathcal{\hat{\mathcal{H}}}_{\mathbf{k}}=\left(\begin{array}{cccccc}
\e_{\Gamma} & -\Delta_{h}(i\sigma_{y}) & -\mathbf{M}_{X}\cdot\boldsymbol{\sigma} & 0 & -\mathbf{M}_{Y}\cdot\boldsymbol{\sigma} & 0\\
\Delta_{h}(i\sigma_{y}) & -\e_{\Gamma} & 0 & \mathbf{M}_{X}\cdot\boldsymbol{\sigma^{*}} & 0 & \mathbf{M}_{Y}\cdot\boldsymbol{\sigma^{*}}\\
-\mathbf{M}_{X}\cdot\boldsymbol{\sigma} & 0 & \e_{X} & -\Delta_{e}(i\sigma_{y}) & 0 & 0\\
0 & \mathbf{M}_{X}\cdot\boldsymbol{\sigma^{*}} & \Delta_{e}(i\sigma_{y}) & -\e_{X} & 0 & 0\\
-\mathbf{M}_{Y}\cdot\boldsymbol{\sigma} & 0 & 0 & 0 & \e_{Y} & -\Delta_{e}(i\sigma_{y})\\
0 & \mathbf{M}_{Y}\cdot\boldsymbol{\sigma^{*}} & 0 & 0 & \Delta_{e}(i\sigma_{y}) & -\e_{Y}
\end{array}\right)
\end{equation}

Following Ref. \cite{RMF}, we integrate out the fermions and expand
for small $\mathbf{M}$, $\Delta$, obtaining the effective action

\begin{eqnarray*}
S_{\mathrm{eff}}\left[\mathbf{M}_{i},\Delta_{i}\right] & = & \frac{2}{(U_{1}+U_{3})}\left(M_{X}^{2}+M_{Y}^{2}\right)-\frac{4}{U_{3}}\Delta_{h}\Delta_{e}\\
 &  & +\frac{1}{2}\int\mathrm{Tr}\left(\hat{G}_{0}\hat{V}\right)^{2}+\frac{1}{2}\int\mathrm{Tr}\left(\hat{G}_{0}\hat{V}\right)^{4}+\mathcal{O}(V^{6})
\end{eqnarray*}
 where $\hat{G}_{0}=\text{diag}(G_{\Gamma},\tilde{G}_{\Gamma},G_{X},\tilde{G}_{X},G_{Y},\tilde{G}_{Y})$
and $\hat{V}$ is the same as $\mathcal{\hat{\mathcal{H}}}$ but with
the diagonal entries set to zero. Here, we introduced the non-interacting
Green's functions. $G_{i,k}^{-1}=i\omega-\e_{i}$ and $\tilde{G}_{i,k}^{-1}\equiv-G_{i,-k}^{-1}=i\omega+\e_{i}$.
To simplify our analysis, we consider the $s^{+-}$ SC gap structure
given by the solution of the linearized gap equations, which gives
$\Delta_{e}/\Delta_{h}=-\frac{1}{\sqrt{2}}$. We obtain:

\begin{eqnarray}
S_{\mathrm{eff}}\left[\mathbf{M}_{i},\Delta_{i}\right] & = & a_{m}\left(M_{X}^{2}+M_{Y}^{2}\right)+a_{s}\Delta^{2}+\frac{u_{s}}{2}\Delta^{4}\nonumber \\
 &  & +\frac{u_{m}^{(1)}+u_{m}^{(2)}}{4}\left(M_{X}^{2}+M_{Y}^{2}\right)^{2}-\frac{u_{m}^{(2)}-u_{m}^{(1)}}{4}\left(M_{X}^{2}-M_{Y}^{2}\right)^{2}+\lambda\left(M_{X}^{2}+M_{Y}^{2}\right)\Delta^{2}\label{S_eff}
\end{eqnarray}
 with $\Delta_{h}\equiv\Delta$ and Ginzburg-Landau coefficients:

\begin{eqnarray}
a_{m} & = & \frac{2}{(U_{1}+U_{3})}+2\int_{k}G_{\Gamma}G_{X}\nonumber \\
a_{s} & = & \frac{4}{\sqrt{2}U_{3}}+2\int_{k}\left(G_{\Gamma}\tilde{G}_{\Gamma}+G_{X}\tilde{G_{X}}\right)\nonumber \\
u_{m}^{(1)} & = & \int_{k}G_{\Gamma}^{2}G_{X}^{2}\nonumber \\
u_{m}^{(2)} & = & \int_{k}G_{\Gamma}^{2}G_{X}G_{Y}\nonumber \\
\lambda & = & 2\int_{k}\left(G_{\Gamma}^{2}\tilde{G}_{\Gamma}G_{X}+\frac{1}{2}G_{X}^{2}\tilde{G}_{X}G_{\Gamma}-\frac{1}{\sqrt{2}}G_{\Gamma}\tilde{G}_{\Gamma}G_{X}\tilde{G}_{X}\right)\nonumber \\
u_{s} & = & 2\int_{k}\left(G_{\Gamma}^{2}\tilde{G}_{\Gamma}^{2}+\frac{1}{2}G_{X}^{2}\tilde{G}_{X}^{2}\right)\label{GL_non_SC}
\end{eqnarray}

Evaluating the momentum integrals above give:

\begin{eqnarray*}
\int_{k}G_{\Gamma}G_{X} & = & -2\pi\rho_{F}T\sum_{n>0}\left\langle \frac{\omega_{n}}{\omega_{n}^{2}+\tilde{\mu}_{X}^{2}}\right\rangle \\
\int_{k}\left(G_{\Gamma}\tilde{G}_{\Gamma}+G_{X}\tilde{G_{X}}\right) & = & -4\pi\rho_{F}T\sum_{n>0}\frac{1}{\omega_{n}}\\
\int_{k}G_{\Gamma}^{2}G_{X}^{2} & = & \pi\rho_{F}T\sum_{n>0}\left\langle \frac{\omega_{n}(\omega_{n}^{2}-3\tilde{\mu}_{X}^{2})}{(\omega_{n}^{2}+\tilde{\mu}_{X}^{2})^{3}}\right\rangle \\
\int_{k}G_{\Gamma}^{2}G_{X}G_{Y} & = & \pi\rho_{F}T\sum_{n>0}\left\langle \frac{\omega_{n}\left[(\omega_{n}^{2}-\tilde{\mu}_{X}\tilde{\mu}_{Y})^{2}-\tilde{\mu}_{X}\tilde{\mu}_{Y}(\tilde{\mu}_{X}+\tilde{\mu}_{Y})^{2}\right]}{(\omega_{n}^{2}+\tilde{\mu}_{X}^{2})^{2}(\omega_{n}^{2}+\tilde{\mu}_{Y}^{2})^{2}}\right\rangle \\
\int_{k}G_{\Gamma}^{2}\tilde{G}_{\Gamma}^{2} & = & \int_{k}G_{X}^{2}\tilde{G}_{X}^{2}=\pi\rho_{F}T\sum_{n>0}\frac{1}{\omega_{n}^{3}}\\
\int_{k}G_{X}^{2}\tilde{G}_{X}G_{\Gamma} & = & \int_{k}G_{\Gamma}^{2}\tilde{G}_{\Gamma}G_{X}=\pi\rho_{F}T\sum_{n>0}\left\langle \frac{\omega_{n}}{(\omega_{n}^{2}+\tilde{\mu}_{X}^{2})^{2}}\right\rangle \\
\int_{k}G_{\Gamma}\tilde{G}_{\Gamma}G_{X}\tilde{G}_{X} & = & \pi\rho_{F}T\sum_{n>0}\left\langle \frac{1}{\omega_{n}(\omega_{n}^{2}+\tilde{\mu}_{X}^{2})}\right\rangle 
\end{eqnarray*}
 where $\rho_{F}$ is the density of states at the Fermi level, $\langle\rangle$
refers to angular averaging over the Fermi surface, and $\tilde{\mu}_{\left(X,Y\right)}=\delta_{0}\pm\delta_{2}\cos2\theta$.

\subsection{Nematicity and superconductivity \label{sec_nematic}}

From Eq.\ref{S_eff}, we can follow the steps in Ref. \cite{RMF}
explained in the main text and introduce the new Hubbard-Stratonovic
fields $\varphi$ and $\psi$ corresponding to $M_{X}^{2}+M_{Y}^{2}$
(thermal fluctuations) and $M_{X}^{2}-M_{Y}^{2}$ (nematic order parameter).
After integrating out the Gaussian magnetic fluctuations in the paramagnetic
phase, the new effective action is

\[
\tilde{S}_{\mathrm{eff}}=\frac{\varphi^{2}}{2g_{m}}-\frac{\psi^{2}}{2u_{m}}+a_{s}\Delta^{2}+\frac{u_{s}}{2}\Delta^{4}+\frac{N}{2}\int_{k}\ln\left[(\psi+\lambda\Delta^{2}+\chi_{0}^{-1})^{2}-\varphi^{2}\right]
\]
where $u_{m}\equiv\frac{u_{m}^{(1)}+u_{m}^{(2)}}{2}$, $g_{m}\equiv\frac{u_{m}^{(2)}-u_{m}^{(1)}}{2}$,
and $N$ is the number of components of the magnetic order parameter.
Here, $\chi_{0}^{-1}\left(\mathbf{Q}_{i}+\mathbf{q},\omega_{n}\right)=a_{m}+q^{2}+f\left(\Omega_{n}\right)$,
where $\Omega_{n}=2\pi nT$ is the Matsubara frequency and $f\left(\Omega_{n}\right)$
is proportional to $\left|\Omega_{n}\right|$ in the normal state
and $\Omega_{n}^{2}$ deep inside the SC state. At the temperature
where the nematic transition meets the SC transition line, we can
restrict our analysis to $\Omega_{n}=0$. Furthermore, since the pnictides
are layered materials, we focus here in the case $d=2$, which gives:

\[
\int\frac{d^{2}q}{\left(2\pi\right)^{2}}\,\ln\left[(r+q^{2})^{2}-\varphi^{2}\right]=\frac{1}{4\pi}\left[2r(1+\ln\Lambda^{2})-(r+\varphi)\ln(r+\varphi)-(r-\varphi)\ln(r-\varphi)\right]
\]
 where $\Lambda$ is the upper momentum cutoff and we defined $r\equiv\psi+\lambda\Delta^{2}+a_{m}$.
To proceed with the saddle point approximation, we rescale $\Delta^{2}\rightarrow n\Delta^{2}$
as well as the quartic coefficients

\begin{eqnarray}
(u_{m},u_{s},g_{m},\lambda) & \rightarrow & \frac{(u_{m},u_{s},g_{m},\lambda)}{n}\label{eq:rescaling}
\end{eqnarray}
 where $n=\frac{NT_{c}}{8\pi}$. Then, $\tilde{S}_{\mathrm{eff}}$
acquires an overall factor of $n$, rendering the saddle-point approximation
exact in the limit $N\rightarrow\infty$. It follows that:

\[
\frac{\tilde{S}_{\mathrm{eff}}}{2g_{m}\, n}=\varphi^{2}-\frac{\psi^{2}}{2u_{m}}+a_{s}\Delta^{2}+\frac{u_{s}}{2}\Delta^{4}+r\,\ln\left(\frac{\Lambda^{4}}{r^{2}-\varphi^{2}}\right)+2r-\varphi\ln\left(\frac{r+\varphi}{r-\varphi}\right)
\]
 where, for convenience, we performed one additional rescaling:

\begin{equation}
\left(u_{m},u_{s},\lambda,a_{m},a_{s},\varphi,\psi,r,\Lambda^{2}\right)\rightarrow2g_{m}\left(u_{m},u_{s},\lambda,a_{m},a_{s},\varphi,\psi,r,\Lambda^{2}\right)\label{eq_rescaling_2}
\end{equation}

Using the saddle-point equation $\frac{\partial\tilde{S}}{\partial\psi}=0$,
we can eliminate $r$, which is given implicitly as a function of
$\varphi$ and $\Delta$:

\begin{equation}
r=\bar{a}_{m}+\lambda\Delta^{2}-u_{m}\ln\left(r^{2}-\varphi^{2}\right)\label{eq_r}
\end{equation}

Furthermore, the cutoff $\Lambda$ has been absorbed into a redifinition
of the quadratic term, $\bar{a}_{m}=a_{m}+2u_{m}\ln\Lambda^{2}$.
The action then can be written as:

\begin{equation}
\bar{S}\equiv\frac{\tilde{S}_{\mathrm{eff}}}{2g_{m}\, n}=\varphi^{2}+\frac{r^{2}}{2u_{m}}+2r-\varphi\ln\left(\frac{r+\varphi}{r-\varphi}\right)+\left(a_{s}-\frac{\lambda a_{m}}{u_{m}}\right)\Delta^{2}+\left(\frac{u_{s}}{2}-\frac{\lambda^{2}}{2u_{m}}\right)\Delta^{4}\label{eq_S_bar}
\end{equation}

Since we are interested in the region of the phase diagram where the
nematic transition line crosses the SC dome, we expand the action
for small $\varphi$ and $\Delta$. In particular, we substitute in
Eq. (\ref{eq_r}):

\begin{equation}
r=r_{0}+b_{1}\varphi^{2}+b_{2}\varphi^{4}+c_{1}\Delta^{2}+c_{2}\Delta^{4}+d\varphi^{2}\Delta^{2}\label{eq:xi}
\end{equation}
 where $r_{0}$ is the solution with $\varphi=0$, $\Delta=0$, and
find the coefficients $b_{i}$, $c_{i}$, and $d$. Substituting this
form in $\bar{S}$ and expanding for small $\varphi$ and $\Delta$,
we obtain:

\[
\bar{S}=a_{n}\varphi^{2}+\frac{u_{n}}{2}\varphi^{4}+\tilde{a}_{s}\Delta^{2}+\frac{\tilde{u}_{s}}{2}\Delta^{4}+\tilde{\lambda}\varphi^{2}\Delta^{2}
\]
 with the Ginzburg-Landau coefficients:

\begin{eqnarray}
\tilde{a}_{s} & = & a_{s}-\frac{\lambda}{u_{m}}\left(a_{m}-r_{0}\right)\nonumber \\
a_{n} & = & 1-\frac{1}{r_{0}}\nonumber \\
\tilde{\lambda} & = & \frac{\lambda}{r_{0}\left(r_{0}+2u_{m}\right)}\nonumber \\
\frac{\tilde{u}_{s}}{2} & = & \frac{u_{s}}{2}-\frac{\lambda^{2}}{r_{0}+2u_{m}}\nonumber \\
\frac{u_{n}}{2} & = & \frac{\left(-r_{0}+u_{m}\right)}{6r_{0}^{3}\left(r_{0}+2u_{m}\right)}
\end{eqnarray}

Since we consider the vicinity of the nematic transition, where $a_{n}=0$,
we can set $r_{0}=1$ in the quartic coefficients $\tilde{\lambda}$,
$\tilde{u}_{s}$, and $u_{n}$. Going back to the original variables
via Eqs. (\ref{eq:rescaling}) and (\ref{eq_rescaling_2}), we then
obtain the results in Eq. (4) of the main text.

\section{Spin-density wave transition inside the SC dome \label{sec:SC}}

To obtain the SDW action deep inside the SC state, where the SC gap
$\Delta$ is nearly saturated, we go back to the original action (\ref{action_formal})
and treat $\Delta$ as a parameter, expanding only in powers of $M_{i}$:

\begin{equation}
S_{\mathrm{eff}}^{(\mathrm{SC})}\left[\mathbf{M}_{i}\right]=a_{m}\left(M_{X}^{2}+M_{Y}^{2}\right)+\frac{u_{m}^{(1)}+u_{m}^{(2)}}{4}\left(M_{X}^{2}+M_{Y}^{2}\right)^{2}-\frac{u_{m}^{(2)}-u_{m}^{(1)}}{4}\left(M_{X}^{2}-M_{Y}^{2}\right)^{2}\label{S_SC}
\end{equation}

As a result of this procedure, the Ginzburg-Landau coefficients depend
now not only on the modified normal Green's function $G_{i,k}$, but
also on the anomalous Green's function $F_{i,k}$:

\begin{eqnarray}
G_{i,k} & = & -\frac{i\omega_{n}+\varepsilon_{\mathbf{k},i}}{\omega_{n}^{2}+\varepsilon_{\mathbf{k},i}^{2}+\Delta_{i}^{2}}\nonumber \\
F_{i,k} & = & \frac{\Delta_{i}}{\omega_{n}^{2}+\varepsilon_{\mathbf{k},i}^{2}+\Delta_{i}^{2}}
\end{eqnarray}

In particular, we obtain:

\begin{eqnarray}
a_{m} & = & \frac{1}{4(U_{1}+U_{3})}+2\int_{k}\left(F_{\Gamma}F_{X}+G_{\Gamma}G_{X}\right)\nonumber \\
u_{m}^{(1)} & = & \int_{k}\left[4F_{\Gamma}F_{X}G_{\Gamma}G_{X}+F_{\Gamma}^{2}\left(F_{X}^{2}+G_{X}\tilde{G}_{X}\right)+F_{X}^{2}G_{\Gamma}\tilde{G}_{\Gamma}+G_{\Gamma}^{2}G_{X}^{2}\right]\nonumber \\
u_{m}^{(2)} & = & \int_{k}\left[F_{\Gamma}^{2}\left(F_{X}F_{Y}+\tilde{G}_{X}G_{Y}\right)+G_{\Gamma}^{2}G_{X}G_{Y}+G_{\Gamma}\tilde{G}_{\Gamma}F_{X}F_{Y}+4F_{\Gamma}F_{X}G_{\Gamma}G_{Y}\right]\label{GL_SC}
\end{eqnarray}
 where $\tilde{G}_{i,k}=-G_{i,-k}$. We define the quasi-particle
excitation energy $E_{i}=\sqrt{\Delta_{i}^{2}+\varepsilon_{i}^{2}}$
and consider the $T=0$ limit, where the Matsubara sum becomes an
integral over frequencies. Performing this integration yields:

\begin{eqnarray}
\int_{k}\left(G_{\Gamma}G_{X}+F_{\Gamma}F_{X}\right) & = & \int_{\mathbf{k}}\,\frac{1}{2(E_{\Gamma}+E_{X})}\left[-1+\frac{\e_{\Gamma}\e_{X}+\D_{\Gamma}\D_{X}}{E_{\Gamma}E_{X}}\right]\label{eq:SC_coeff}\\
\int_{k}G_{\Gamma}G_{X}F_{\Gamma}F_{X} & = & \int_{\mathbf{k}}\,\frac{1}{4(E_{\Gamma}+E_{X})^{3}}\left[-\frac{\D_{\Gamma}\D_{X}}{E_{\Gamma}E_{X}}+\frac{\D_{\Gamma}\D_{X}\e_{\Gamma}\e_{X}}{(E_{\Gamma}E_{X})^{2}}\Xi\right]\\
\int_{k}F_{\Gamma}^{2}F_{X}^{2} & = & \int_{\mathbf{k}}\,\frac{1}{4(E_{\Gamma}+E_{X})^{3}}\left(\frac{\D_{\Gamma}\D_{X}}{E_{\Gamma}E_{X}}\right)^{2}\Xi\\
\int_{k}\left(F_{X}^{2}G_{\Gamma}\tilde{G}_{\Gamma}+F_{\Gamma}^{2}G_{X}\tilde{G}_{X}\right) & = & \int_{\mathbf{k}}\,\frac{1}{4(E_{\Gamma}+E_{X})^{3}}\left[-\frac{\D_{\Gamma}^{2}+\D_{X}^{2}}{E_{\Gamma}E_{X}}-\frac{\D_{\Gamma}^{2}\e_{X}^{2}+\D_{X}^{2}\e_{\Gamma}^{2}}{(E_{\Gamma}E_{X})^{2}}\Xi\right]\\
\int_{k}G_{\Gamma}^{2}G_{X}^{2} & = & \int_{\mathbf{k}}\,\frac{1}{4(E_{\Gamma}+E_{X})^{3}}\left[1-\frac{\e_{\Gamma}^{2}+\e_{X}^{2}+4\e_{\Gamma}\e_{X}}{E_{\Gamma}E_{X}}+\left(\frac{\e_{X}\e_{\Gamma}}{E_{\Gamma}E_{X}}\right)^{2}\Xi\right]
\end{eqnarray}
 where $\Xi=3+\frac{E_{\Gamma}^{2}+E_{X}^{2}}{E_{\Gamma}E_{X}}$.
We also obtain:

\begin{eqnarray}
\int_{k}F_{\Gamma}^{2}F_{X}F_{Y} & = & \int_{\mathbf{k}}\frac{\D_{\Gamma}^{2}\D_{X}^{2}}{E_{\Gamma}^{2}}\frac{\mathcal{A}+\mathcal{B}}{\mathcal{D}}\label{eq:SC_coeff_cont}\\
\int_{k}F_{\Gamma}^{2}\tilde{G}_{X}G_{Y} & = & \int_{\mathbf{k}}-\D_{\Gamma}^{2}\frac{\mathcal{C}+\mathcal{F}}{\mathcal{D}}-\frac{\D_{\Gamma}^{2}\e_{X}\e_{Y}}{E_{\Gamma}^{2}}\frac{\mathcal{A}+\mathcal{B}}{\mathcal{D}}\\
\int_{k}F_{X}F_{Y}\tilde{G}_{\Gamma}G_{\Gamma} & = & \int_{\mathbf{k}}-\D_{X}^{2}\frac{\mathcal{C}+\mathcal{F}}{\mathcal{D}}-\frac{\D_{X}^{2}\e_{\Gamma}^{2}}{E_{\Gamma}^{2}}\frac{\mathcal{A}+\mathcal{B}}{\mathcal{D}}\\
\int_{k}G_{\Gamma}G_{Y}F_{\Gamma}F_{X} & = & \int_{\mathbf{k}}-\D_{X}\D_{\Gamma}\frac{\mathcal{C}+\mathcal{F}}{\mathcal{D}}+\frac{\D_{\Gamma}\D_{X}\e_{\Gamma}\e_{Y}}{E_{\Gamma}^{2}}\frac{\mathcal{A}+\mathcal{B}}{\mathcal{D}}\\
\int_{k}G_{\Gamma}^{2}G_{X}G_{Y} & = & \int_{\mathbf{k}}\frac{1}{2E_{X}E_{Y}(E_{X}+E_{Y})}-(\e_{\Gamma}^{2}+2\e_{\Gamma}(\e_{X}+\e_{Y})+\e_{X}\e_{Y}+2E_{\Gamma}^{2})\frac{\mathcal{C}+\mathcal{F}}{\mathcal{D}}+\frac{\e_{\Gamma}^{2}\e_{X}\e_{Y}-E_{\Gamma}^{4}}{E_{\Gamma}^{2}}\frac{\mathcal{A}+\mathcal{B}}{\mathcal{D}}\nonumber \\
\end{eqnarray}
 with:

\begin{eqnarray}
\mathcal{A} & = & E_{\Gamma}(E_{\Gamma}+E_{X}+E_{Y})(2E_{\Gamma}+E_{X}+E_{Y})\nonumber \\
\mathcal{B} & = & (E_{\Gamma}+E_{X})(E_{\Gamma}+E_{Y})(E_{X}+E_{Y})\nonumber \\
\mathcal{C} & = & (E_{\Gamma}+E_{X}+E_{Y})(-E_{\Gamma}^{2}+E_{X}E_{Y})\nonumber \\
\mathcal{F} & = & E_{\Gamma}(E_{\Gamma}+E_{X})(E_{c}+E_{Y})\nonumber \\
\mathcal{D} & = & 4E_{\Gamma}E_{X}E_{Y}(E_{\Gamma}+E_{X})^{2}(E_{\Gamma}+E_{Y})^{2}(E_{X}+E_{Y})\nonumber \\
\end{eqnarray}

Following Ref. \cite{S_Maiti10}, we set $\Delta_{\Gamma}=\Delta$ and
$\Delta_{X}=-\Delta/\sqrt{2}$. The nature of the mean-field SDW transition
inside the SC dome is determined by the quartic coefficients $u_{m}\equiv\frac{u_{m}^{(1)}+u_{m}^{(2)}}{2}$,
$g_{m}\equiv\frac{u_{m}^{(2)}-u_{m}^{(1)}}{2}$. In the absence of
SC, $u_{m}<0$ always at $T=0$, regardless of the band structure
parameters $\delta_{0}$ and $\delta_{2}$, implying that at $T=0$
the SDW transition is first-order. This is illustrated in Fig. \ref{fig:1}a,
where we plot the $T=0$ value of $u_{m}$ given by Eq. (\ref{GL_non_SC})
- i.e. without SC - as function of $\delta_{0}/\varepsilon_{F}$ ($\varepsilon_{F}$
is the fermi energy) for $\delta_{2}=0$. The presence of SC can significantly
change this result: in Fig. \ref{fig:1}b, we plot $u_{m}$ at $T=0$
in the presence of SC, as given by Eq. (\ref{GL_SC}), as function
of $\delta_{0}/\Delta$ for $\delta_{2}=0$. Clearly, for a large
enough SC gap $\Delta$, $u_{m}$ changes sign, yielding a second-order
SDW transition at $T=0$ as long as $g_{m}<u_{m}$ as well. In Fig.
2 of the main text, we plot the $T=0$ value of the ratio $\alpha=u_{m}/g_{m}$
in the entire $\left(\delta_{0},\delta_{2}\right)$ parameter space.

\begin{figure}[htp]
\centering{}\includegraphics[width=0.5\columnwidth]{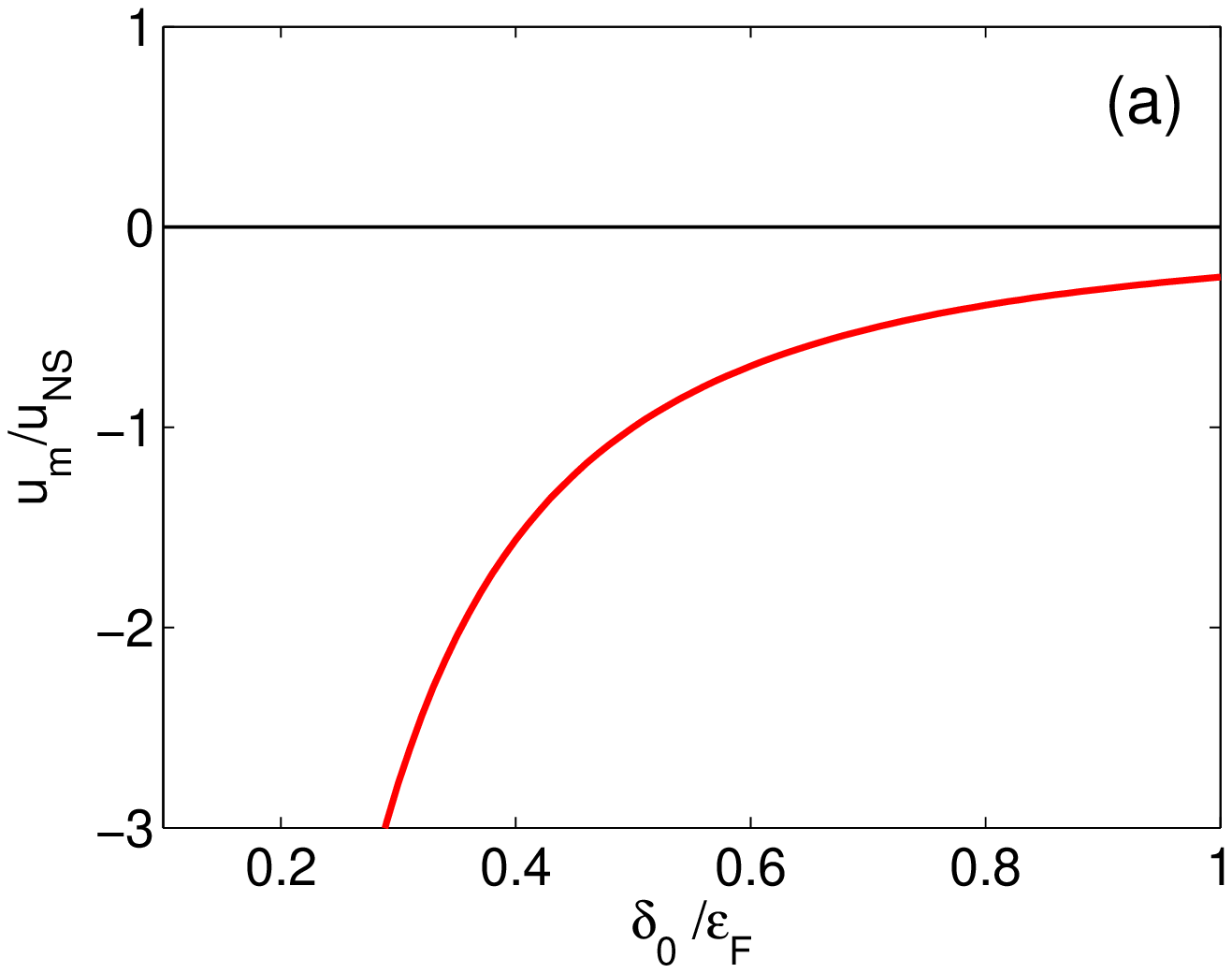}\hfill{}\includegraphics[width=0.5\columnwidth]{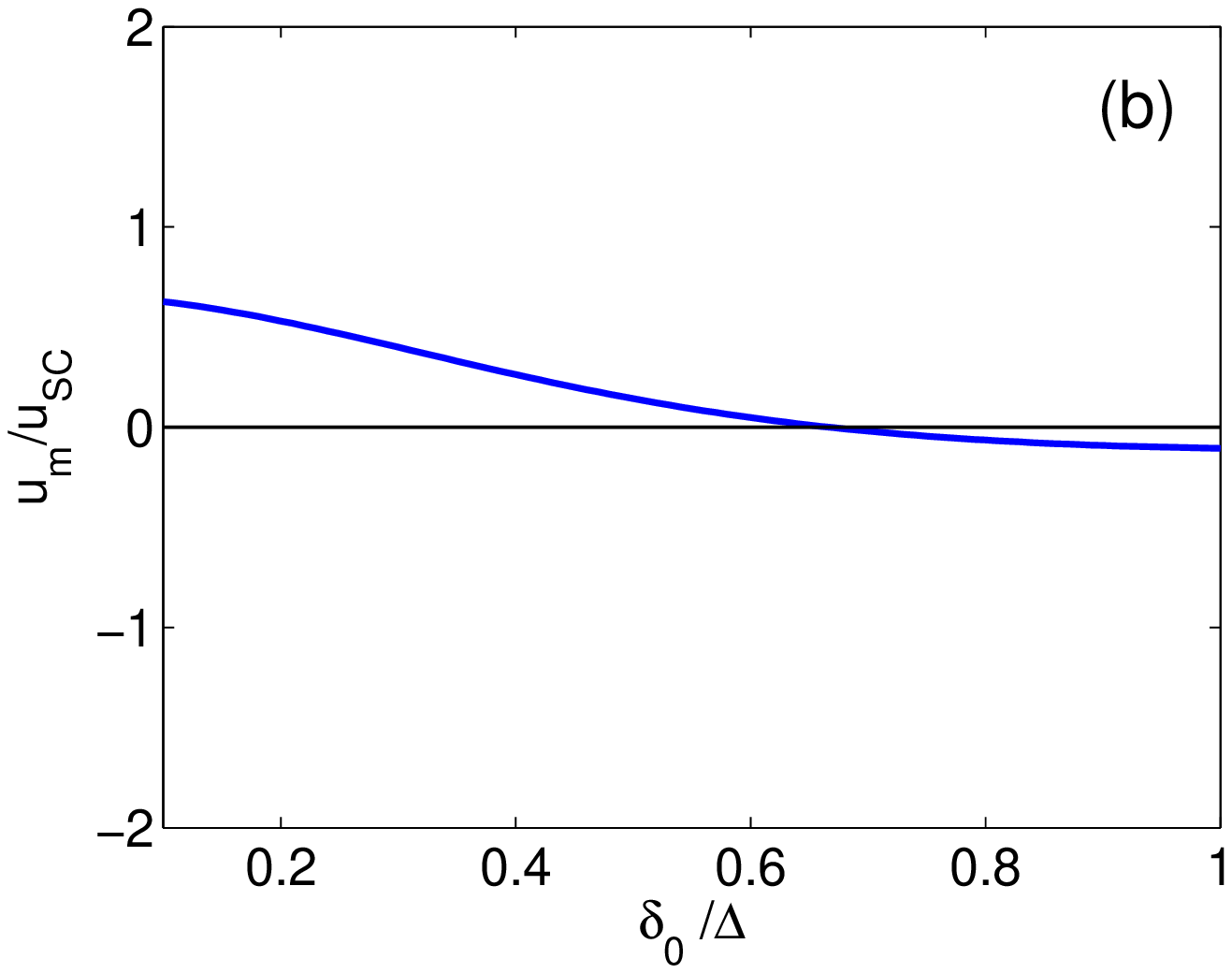}
\caption{\label{fig:1} Plot of the $T=0$ coefficient $u_{m}$ (a) in the
absence of SC (in units of $u_{NS}\equiv\frac{\rho_{F}}{\varepsilon_{F}^{2}}$,
where $\varepsilon_{F}$ is the fermi energy) and (b) in the presence
of SC (in units of $u_{SC}\equiv\frac{\rho_{F}}{\Delta^{2}}$) as
a function of the band dispersion parameter $\delta_{0}$. We set
$\delta_{2}=0$ in these plots.}
\end{figure}

\section{Nematic transition inside the SC dome}

By using the magnetic action inside the SC state, Eq. (\ref{S_SC}),
with Ginzburg-Landau coefficients renormalized by the SC order, we
can go beyond mean-field to study the nematic transition temperature
near $T=0$. In the main text, we showed that at $T=0$ and $d=2$,
$z=1$, there is a simultaneous first-order SDW/nematic transition.
We now extend this analysis to finite temperatures. Repeating the
same procedure as described in Section \ref{sec_nematic}, we introduce
the Hubbard-Stratonovich fields $\varphi$ and $\psi$ and obtain
the saddle-point equations:

\begin{eqnarray}
\frac{\varphi}{\bar{g}_{m}} & = & 2T\sum_{n}\int q\, dq\left(\frac{1}{\psi+a_{m}+q^{2}+\Omega_{n}^{2}-\varphi}-\frac{1}{\psi+a_{m}+q^{2}+\Omega_{n}^{2}+\varphi}\right)\nonumber \\
\frac{\psi}{\bar{u}_{m}} & = & 2T\sum_{n}\int q\, dq\left(\frac{1}{\psi+a_{m}+q^{2}+\Omega_{n}^{2}-\varphi}+\frac{1}{\psi+a_{m}+q^{2}+\Omega_{n}^{2}+\varphi}\right)
\end{eqnarray}
where $\left(\bar{g}_{m},\bar{u}_{m}\right)=\frac{N}{8\pi}\left(g_{m},u_{m}\right)$.
To perform the Matsubara frequency summation, we use the identity
\begin{equation}
T\sum_{n}f(i\Omega_{n})=\int_{-\infty}^{\infty}\frac{dx}{2\pi}\coth\left(\frac{x}{2T}\right)\text{Im}[f(x)]\label{eq:bosonsum}
\end{equation}
 valid for any function $f$ of bosonic Matsubara frequencies $\Omega_{n}=2n\pi T$.
After defining $r=a_{m}+\psi$ and the rescaled quantities $(\varphi,r)=(\bar{\varphi},\bar{r})\bar{g}_{m}^{2}$,
$T=\bar{T}\bar{g}_{m}$ ,we obtain the saddle-point equations:

\begin{eqnarray}
\bar{\varphi} & = & 2\bar{T}\left(\ln\,\sinh\frac{\sqrt{\bar{r}+\bar{\phi}}}{2\bar{T}}-\ln\,\sinh\frac{\sqrt{\bar{r}-\bar{\phi}}}{2\bar{T}}\right)\\
\bar{r} & = & \bar{a}_{m}-2\bar{T}\alpha\left(\ln\,\sinh\frac{\sqrt{\bar{r}+\bar{\phi}}}{2\bar{T}}+\ln\,\sinh\frac{\sqrt{\bar{r}-\bar{\phi}}}{2\bar{T}}\right)
\end{eqnarray}
where $\bar{a}_{m}=\left(a_{m}+2\Lambda_{q}\bar{u}_{m}\right)/\bar{g}_{m}^{2}$,
$\Lambda_{q}$ is the momentum integration upper cutoff, and $\alpha=u_{m}/g_{m}$.
In Fig. \ref{fig:2}, we plot the renormalized control parameter $\bar{a}_{m}$
as function of $\bar{\varphi}$ for different fixed temperatures.
We see that for small $\bar{T}$ the first instability (corresponding
to the largest $\bar{a}_{m}$) is at $0<\bar{\varphi}<\bar{r}$, indicating
a first-order nematic transition. As $\bar{T}$ increases, the first
instability moves towards $\bar{\varphi}=0$, indicating a second-order
nematic transition. Because here we considered $d=2$, the magnetic
transition only happens at zero temperature, but for an anisotropic
quasi-2D system, following Ref. \cite{RMF}, the magnetic transition
will remain first-order and simultaneous to the nematic transition
up to a temperature $T_{\mathrm{merge}}$, above which one of the
two transition become second-order and split from the other.

\begin{figure}[htp]
\centering{}\includegraphics[width=0.6\columnwidth]{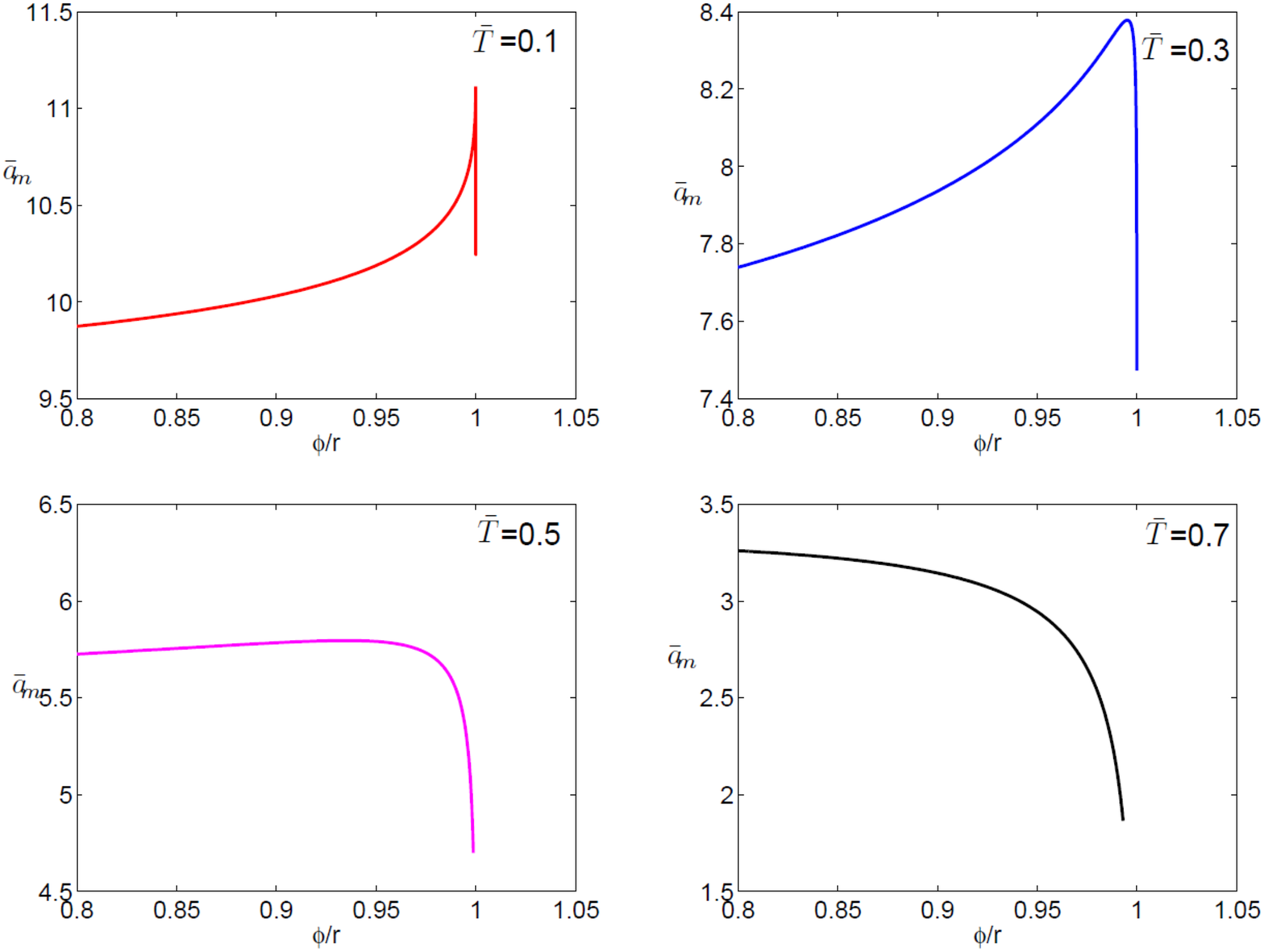} \caption{\label{fig:2} Plot of $\bar{a}_{m}$ as a function of $\varphi/r=\bar{\varphi}/\bar{r}$
inside the SC dome for different temperatures. }
\end{figure}

\section{On the peculiarities of the Hubbard-Stratonovich transformation}

In this section we discuss one subtle issue related to the regularization
of the integrals in the Hubbbard-Stratonovich transformation. Consider
for instance the transformation from an effective action for the SDW
fields $\Delta_{X}$ and $\Delta_{Y}$ deep inside the SC state, Eq.
(\ref{S_SC}). In the notation we introduced after Eq. (\ref{S_SC}),
\begin{equation}
S_{\mathrm{eff}}\left[\mathbf{M}_{i}\right]=a_{m}\left(M_{X}^{2}+M_{Y}^{2}\right)+\frac{u_{m}}{2}\left(M_{X}^{2}+M_{Y}^{2}\right)^{2}-\frac{g_{m}}{2}\left(M_{X}^{2}-M_{Y}^{2}\right)^{2}\label{S_SC2}
\end{equation}

As we discussed above, in the SC state $u_{m}$ and $g_{m}$ are both
positive. The partition function, from which we extract the free energy,
is given by $Z=\int dM_{X}dM_{Y}e^{-S_{\mathrm{eff}}\left[\mathbf{M}_{i}\right]}$.
The rationale behind the Hubbard-Stratonovich transformation is to
rewrite the partition function as an integral over the new fields
$\psi$ and $\varphi$ which describe the fluctuations of $M_{X}^{2}+M_{Y}^{2}$
and $M_{X}^{2}-M_{Y}^{2}$, respectively. This is done by expressing
the quartic terms in $e^{-S_{\mathrm{eff}}\left[\mathbf{M}_{i}\right]}$
as integrals over the new fields $\psi$ and $\varphi$ of some new
effective action $S_{\mathrm{eff}}(M_{i},\psi,\varphi)$ which depends
only linearly on $M_{X}^{2}+M_{Y}^{2}$ and $M_{X}^{2}-M_{Y}^{2}$.
The integrals over $M_{X}$ and $M_{Y}$ in the partition function
can then be easily evaluated. Exponentiating the result one expresses
the partition function as $Z=\int d\psi d\varphi e^{-S_{\mathrm{eff}}\left[\psi,\varphi\right]}$.

For the nematic field $\varphi$, the computational procedure is free
from subtleties. We use the mathematical identity which states that
$e^{\frac{ax^{2}}{2}}$ can be expressed, for positive $a$, as 
\begin{equation}
\mathrm{e}^{\frac{ax^{2}}{2}}=\frac{1}{\sqrt{2\pi a}}\int_{-\infty}^{\infty}dy\,\mathrm{e}^{\frac{-y^{2}}{2a}+xy}\label{chu_1}
\end{equation}

Indeed, 
\begin{equation}
\int_{-\infty}^{\infty}dy\,\mathrm{e}^{\frac{-y^{2}}{2a}+xy}=\int_{-\infty}^{\infty}dy\,\mathrm{e}^{\frac{-(y-ax)^{2}}{2a}}\mathrm{e}^{\frac{ax^{2}}{2}},\label{chu_2}
\end{equation}
 and the integral over $y$ converges and can be trivially evaluated
by shifting variables: 
\begin{equation}
\int_{-\infty}^{\infty}dy\,\mathrm{e}^{\frac{-(y-ax)^{2}}{2a}}=2\int_{0}^{\infty}dy\,\mathrm{e}^{-y^{2}/(2a)}=\sqrt{2\pi a}\label{chu_9}
\end{equation}

Applying this transformation to the $g_{m}$ term in $e^{-S_{eff}^{SC}\left[\mathbf{M}_{i}\right]}$,
we obtain:
\begin{equation}
\mathrm{e}^{\frac{g_{m}}{2}\left(M_{X}^{2}-M_{Y}^{2}\right)^{2}}=\frac{1}{\sqrt{2\pi g_{m}}}\int_{-\infty}^{\infty}d\varphi\,\mathrm{e}^{\frac{-\varphi^{2}}{2g_{m}}+\varphi\left(M_{X}^{2}-M_{Y}^{2}\right)}\label{chu_3}
\end{equation}

For the $u_{m}$ term, however, the Hubbard-Stratonovich transformation
is trickier because the corresponding term in the effective action
is 
\begin{equation}
\mathrm{e}^{-\frac{u_{m}}{2}\left(M_{X}^{2}+M_{Y}^{2}\right)^{2}}\label{chu_4}
\end{equation}
where, as we said, $u_{m}>0$. We can still formally apply the Hubbard-Stratonovich
transformation and obtain 
\begin{equation}
\mathrm{e}^{-\frac{u_{m}}{2}\left(M_{X}^{2}+M_{Y}^{2}\right)^{2}}=\frac{1}{\Lambda}\int_{-\infty}^{\infty}d\psi\,\mathrm{e}^{\frac{\psi^{2}}{2u_{m}}+\psi\left(M_{X}^{2}+M_{Y}^{2}\right)}\label{chu_5}
\end{equation}
but now the normalization factor

\begin{equation}
\Lambda=2\int_{0}^{\infty}dy\, e^{\frac{y^{2}}{2u_{m}}}\label{chu_6}
\end{equation}
 diverges.

One can avoid the divergence by introducing the imaginary field $\psi\to i{\tilde{\psi}}$
instead of the real one, i.e., by writing the exact and well-defined
relation 
\begin{equation}
e^{-\frac{u_{m}}{2}\left(M_{X}^{2}-M_{Y}^{2}\right)^{2}}=\frac{1}{\sqrt{2\pi u_{m}}}\int_{-\infty}^{\infty}d\psi\,\mathrm{e}^{\frac{-{\tilde{\psi}}^{2}}{2u_{m}}+i{\tilde{\psi}}\left(M_{X}^{2}+M_{Y}^{2}\right)}\label{chu_7}
\end{equation}

However, later in the calculation the effective action is analyzed
as a function of $\tilde{\psi}$ in the complex plane\cite{subir}
and is taken at the position of the extreme along the \textit{imaginary}
axis of $\tilde{\psi}$, i.e., one actually returns to real $\psi$,
for which the normalization factor in Eq. (\ref{chu_5}) is the issue.

We argue that the more appropriate way to proceed is to consider $\Lambda$
in Eq. (\ref{chu_6}) as the limit

\begin{equation}
\Lambda=\lim_{\delta\rightarrow\pi}\Lambda_{\delta}\label{chu_8}
\end{equation}
 where 
\begin{equation}
\Lambda_{\delta}=2\int_{0}^{\infty}dy\,\mathrm{e}^{\frac{-y^{2}e^{i\delta}}{2u_{m}}}\label{chu_10}
\end{equation}
 The well-behaved integral in (\ref{chu_9}) corresponds to $\delta=0$.

\begin{figure}[htp]
\begin{centering}
\includegraphics[width=0.6\columnwidth]{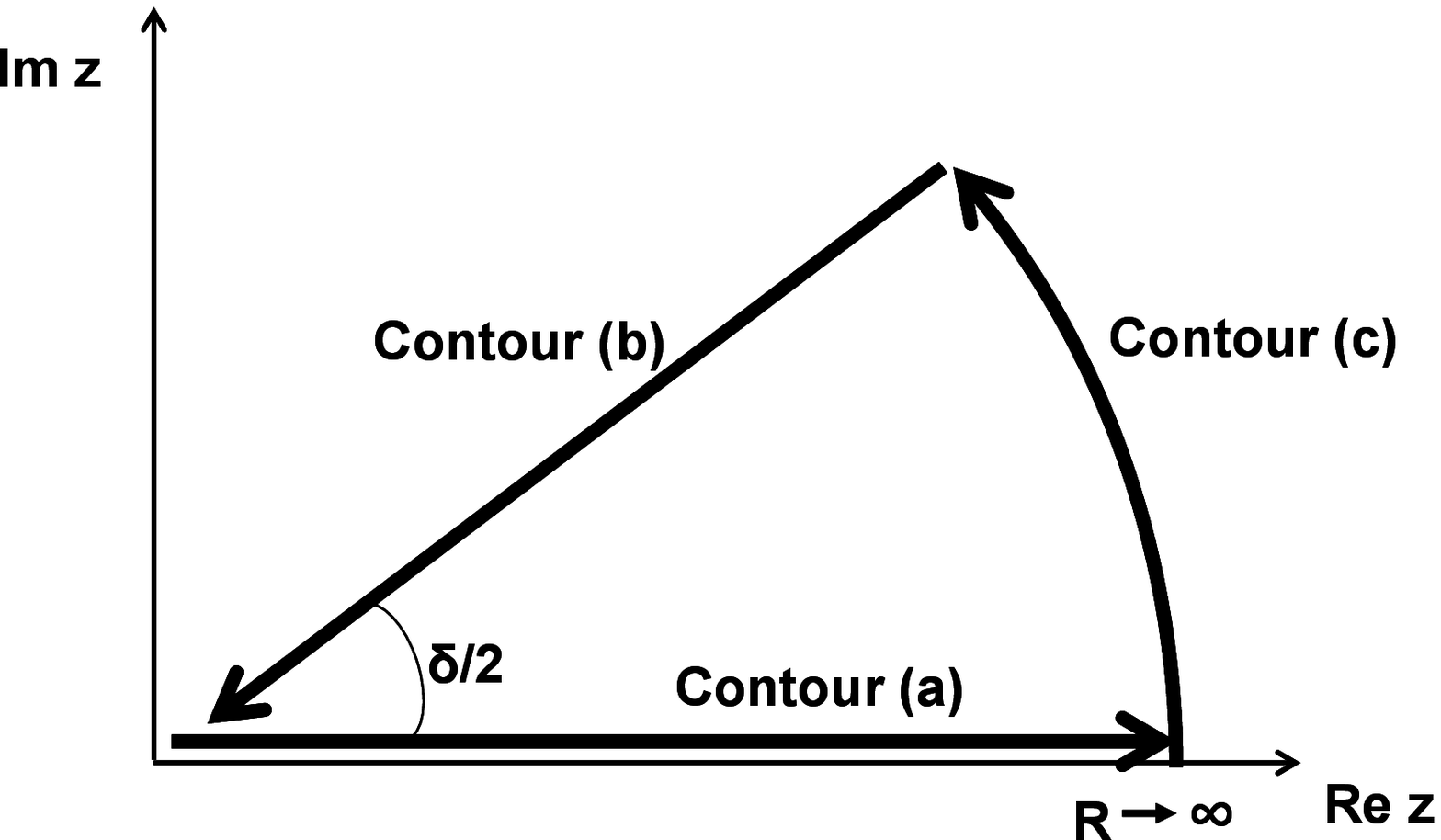}
\par\end{centering}

\caption{Integration contour for the evaluation of the integral $\Lambda_{\delta}$
from Eq. (\ref{chu_10}). The radius $R$ has to be set to infinity
at the end of the calculation. \label{fig:contour}}
\end{figure}

To evaluate $\Lambda_{\delta}$, consider the integral $J_{\delta}=\oint dze^{\frac{-z^{2}e^{i\delta}}{2}}$
over a complex variable $z$, taken over the contour shown in Fig.
\ref{fig:contour}. The contour consists of two lines - one along
the positive real axis and one along a line in the upper half-plane
of $z$, directed at an angle $\delta/2$ with respect to the real
axis - and the arc with radius $R$ which will be set to infinity
at the end of the calculation. We label the corresponding contributions
as $J_{a},J_{b}$, and $\frac{J_{c}}{2}$, respectively: 
\begin{eqnarray}
J_{a} & = & \int_{0}^{R}\mathrm{e}^{-y^{2}/2u_{m}}dy\nonumber \\
J_{b} & = & -\mathrm{e}^{i\delta/2}\int_{0}^{R}dy\,\mathrm{e}^{-y^{2}\mathrm{e}^{i\delta}/(2u_{m})}\nonumber \\
\frac{J_{c}}{2} & = & iR\int_{0}^{\delta/2}\mathrm{e}^{i\theta}\mathrm{e}^{-R^{2}\mathrm{e}^{2i\theta}/(2u_{m})}d\theta\label{chu_11}
\end{eqnarray}

The integral $J_{a}$ is elementary and in the limit of $R\to\infty$
yields $\sqrt{\pi u_{m}/2}$. The second integral at $R\to\infty$
becomes $J_{b}\to-(1/2)\mathrm{e}^{i\delta/2}\Lambda_{\delta}$, and
the integral $J_{c}$ over the arc needs to be carefully analyzed.

\begin{figure}[htp]
\centering{}\includegraphics[width=0.4\columnwidth]{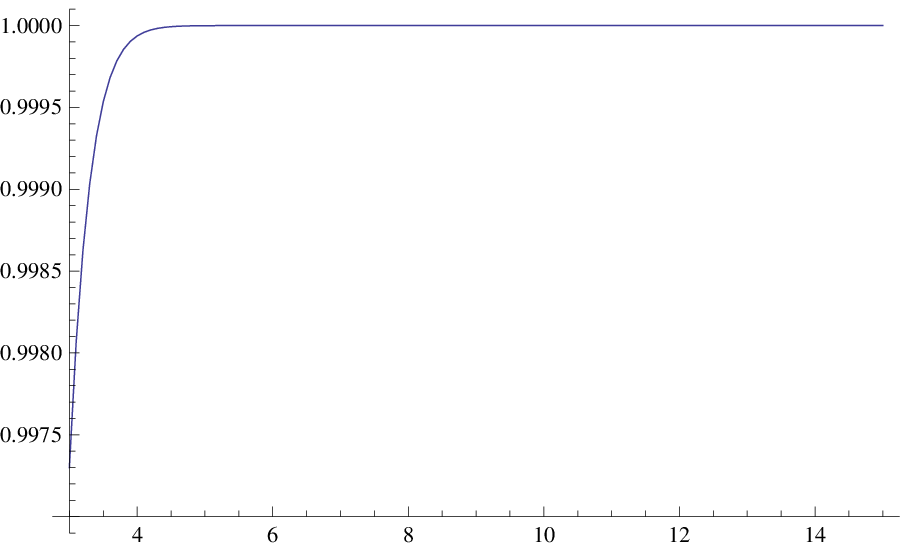}\hfill{}\includegraphics[width=0.4\columnwidth]{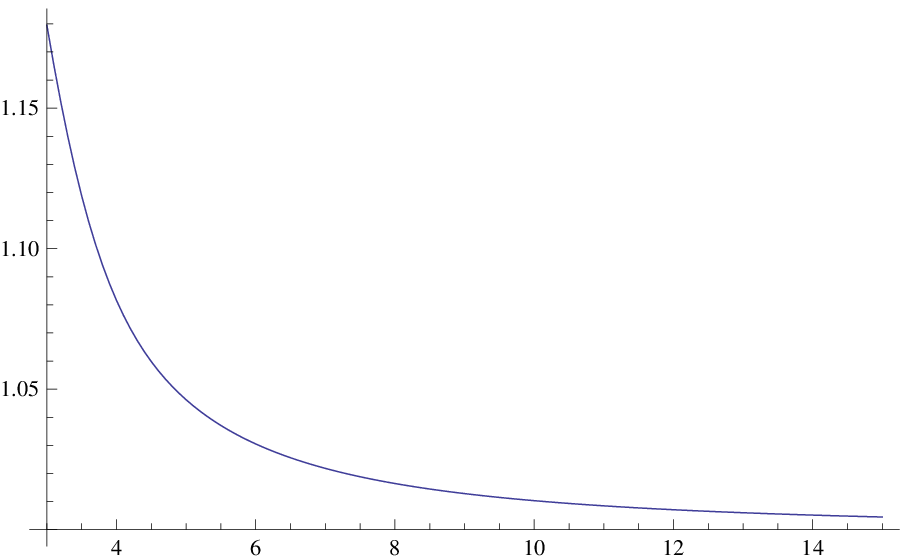}
\caption{Left and right panels - real and imaginary parts of the integral over
the arc of the contour shown in Fig. \ref{fig:contour}, for $\theta$
exactly equal to $\pi/2$, plotted as function of the radius of the
arc $R$. $\mathrm{Re}J_{c}$ rapidly approaches $-\sqrt{2\pi u_{m}}$
while $\mathrm{Im}J_{c}$ approaches $(2u_{m}/R)\mathrm{e}^{R^{2}/(2u_{m})}$.
Neither of the integrals oscillates at large $R$. The vertical axes
are $-\mathrm{Re}J_{c}/\sqrt{2\pi u_{m}}$ and $\mathrm{Im}J_{c}(R/2u_{m})\mathrm{e}^{-R^{2}/(2u_{m})}$,
respectively. \label{fig_J_c_pi}}
\end{figure}

Because the function $\mathrm{e}^{\frac{-z^{2}e^{i\delta}}{2}}$ is
analytic for any finite $z$, we have $J=J_{a}+J_{b}+J_{c}=0$, hence
\begin{equation}
\Lambda_{\delta}=\mathrm{e}^{-i\delta/2}\left(\sqrt{2\pi u_{m}}+J_{c}\right)\label{chu_12}
\end{equation}

\begin{figure}[htp]
\centering{}\includegraphics[width=0.4\columnwidth]{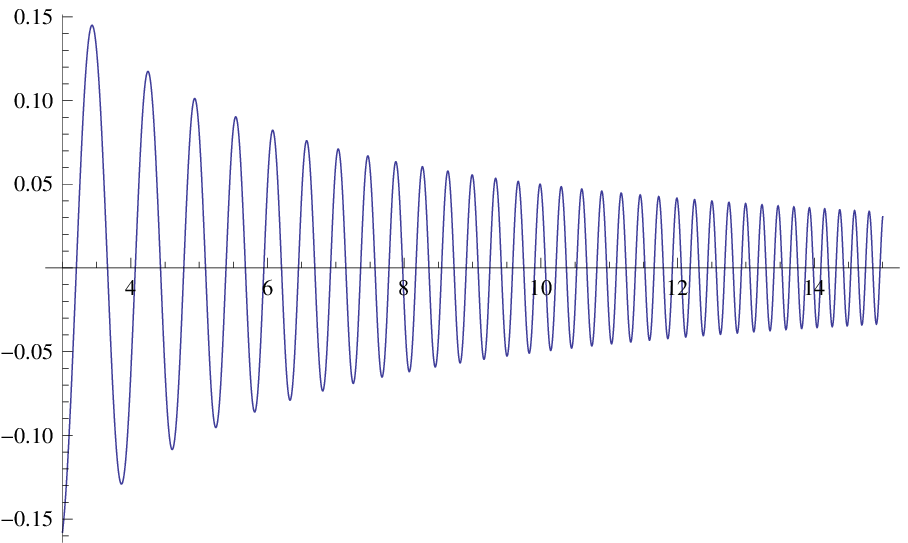}\hfill{}\includegraphics[width=0.4\columnwidth]{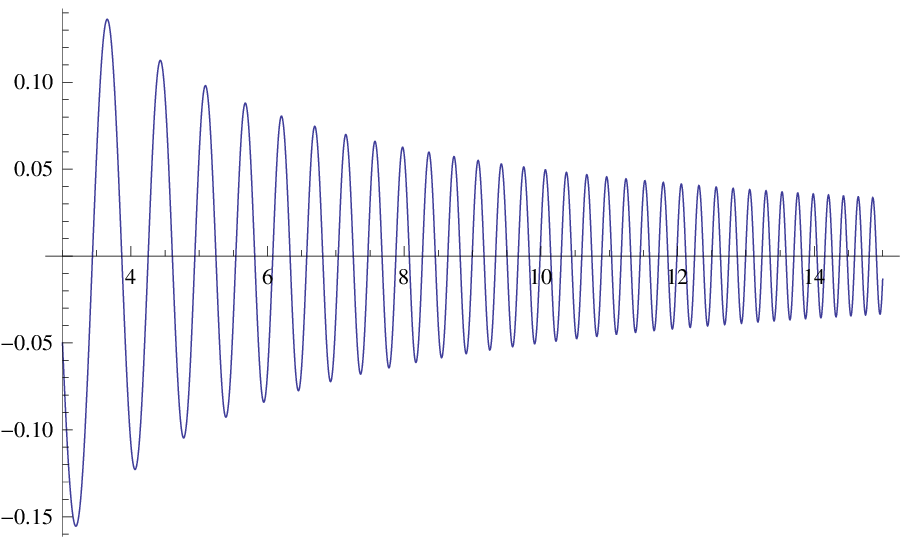}
\caption{The same as in Fig.\ref{fig_J_c_pi} but for $\theta=\pi/4$. Both
$\mathrm{Re}J_{c}$ and $\mathrm{Im}J_{c}$ rapidly oscillate with
$R$. We verified that oscillations persist for all $\theta<\pi/2$.
The value of $R$ at which oscillations begin gets progressively larger
as $\theta$ increases and diverges at $\theta\to\pi/2$ \label{fig_J_c_pi4}}
\end{figure}

The issue then is to calculate $J_{c}$. It is a complex function
$J_{c}=\mathrm{Re}J_{c}+i\,\mathrm{Im}J_{c}$, whose real and imaginary
parts are given by 
\begin{eqnarray}
 &  & \mathrm{Re}J_{c}=-2R\int_{0}^{\delta/2}\mathrm{e}^{-(R^{2}\cos{2\theta})/(2u_{m})}\sin(\theta-R^{2}\sin{2\theta})\, d\theta\nonumber \\
 &  & \mathrm{Im}J_{c}=2R\int_{0}^{\delta/2}\mathrm{e}^{-(R^{2}\cos{2\theta})/(2u_{m})}\cos(\theta-R^{2}\sin{2\theta})\, d\theta\label{chu_14}
\end{eqnarray}

At $\delta$ exactly equal to $\pi$, both terms can be readily calculated
numerically. We show the results in Fig.\ref{fig_J_c_pi}. While the
real part $\mathrm{Re}J_{c}$ rapidly approaches $-\sqrt{2\pi u_{m}}$,
the imaginary part $\mathrm{Im}J_{c}$ rapidly approaches $(2u_{m}/R)\mathrm{e}^{R^{2}/(2u_{m})}$.
As a result, $\Lambda_{\pi}\approx(2u_{m}/R)\mathrm{e}^{R^{2}/(2u_{m})}$,
which is nothing but $2\int_{0}^{R}dy\,\mathrm{e}^{y^{2}/(2u_{m})}$.
Obviously, $\Lambda_{\pi}$ diverges when $R\to\infty$.

We found that the behavior of $\mathrm{Re}J_{c}$ and $\mathrm{Im}J_{c}$
changes qualitatively at large $R$ once $\delta$ becomes different
than $\pi$ (see Fig. \ref{fig_J_c_pi4}). In particular, starting
from some critical $R_{0}$, both $\mathrm{Re}J_{c}$ and $\mathrm{Im}J_{c}$
become oscillating functions of $R$. As a result, there is an infinite
set of $R_{i}$ at which $\mathrm{Re}J_{c}=0$ and an infinite set
of $R_{j}$ at which $\mathrm{Im}J_{c}=0$. We show this behavior
in Fig. \ref{fig_J_c_pi4} for $\delta=\pi/2$. The value $R_{0}$
at which oscillations begin is infinite at $\delta=\pi$, but is finite
at any $\delta<\pi$ and its value decreases as $\pi-\delta$ increases.

Because both $\mathrm{Re}J_{c}$ and $\mathrm{Im}J_{c}$ oscillate,
the limit of each of these functions at $R\to\infty$ depends on how
one approaches $R\to\infty$. In particular, one can use the fact
that there is an infinite set of $R$'s for which $\mathrm{Re}J_{c}=0$,
$\left\{ R_{i}\right\} $, and another one for which $\mathrm{Im}J_{c}=0$,
$\left\{ R_{j}\right\} $. By approaching the limit $R\to\infty$
via the sets $\left\{ R_{i}\right\} $ and $\left\{ R_{j}\right\} $,
we obtain $\lim\limits _{R\rightarrow\infty}J_{c}=0$. We emphasize
that this is only possible for $\delta<\pi$, when both $\mathrm{Re}J_{c}$
and $\mathrm{Im}J_{c}$ oscillate with $R$. Substituting $J_{c}=0$
into (\ref{chu_12}) we find 
\begin{equation}
\Lambda_{\delta}=\mathrm{e}^{-i\delta/2}\sqrt{2\pi u_{m}}\label{chu_15}
\end{equation}

Hence, $\Lambda_{\pi}=\lim\limits _{\delta\rightarrow\pi}\Lambda_{\delta}=-i\sqrt{2\pi u_{m}}$,
i.e. it is finite. Note that Eq. (\ref{chu_15}) gives the same result
for $\Lambda_{\delta}$ which we would obtain by formally substituting
$u_{m}\to u_{m}\mathrm{e}^{-i\delta}$ into Eq. (\ref{chu_9}). For
other ways to regularize such an integral see Refs. \cite{ref1,ref2}.

The same computational scheme can be applied to the evaluation of
the action near the extremum of $S_{\mathrm{eff}}\left[\psi,\varphi\right]$.
Taken as a function of $\psi$, the extremum is a maximum rather than
a minimum. Expanding the action near the maximum at $\psi=\psi_{0}$,
we get $S_{\mathrm{eff}}\left[\psi,\varphi\right]=S_{\mathrm{eff}}\left[\psi_{0},\varphi\right]-A(\psi-\psi_{0})^{2}$
with $A>0$. For the partition function, we then obtain 
\begin{equation}
Z=Z_{0}\int d\psi\,\mathrm{e}^{A(\psi-\psi_{0})^{2}}\label{chu_16}
\end{equation}
 Formally, the integral diverges and makes the expansion near $\psi_{0}$
problematic. However, once we define the integral over $\psi$ in
the same way as above, the integral becomes finite and one can apply
a conventional reasoning (e.g., large N expansion~\cite{subir})
to justify the approximation in which $S_{\mathrm{eff}}\left[\psi,\varphi\right]$,
viewed as a function of $\psi$, is taken at the maximum.

\end{widetext}


\begin{thebibliography}{10}
\bibitem{Sachdev10} S. Sachdev, \emph{Quantum phase transitions of
antiferromagnets and the cuprate superconductors}, Lecture Notes in
Physics v. \textbf{843}, Springer, Berlin (2012).

\bibitem{Wolfle11} P. W\"olfle and E. Abrahams, Phys. Rev. B \textbf{84},
041101(R) (2011).

\bibitem{Wolfle_RMP} H. v. L\"ohneysen, A. Rosch, M. Vojta, and
P. W\"olfle, Rev. Mod. Phys. \textbf{79}, 1015 (2007).

\bibitem{Moon10} E. G. Moon and S. Sachdev, Phys. Rev. B \textbf{82},
104516 (2010); Phys. Rev. B. \textbf{85}, 184511 (2012).

\bibitem{reviews} K. Ishida, Y. Nakai, and H. Hosono, J. Phys. Soc.
Jpn. \textbf{78}, 062001 (2009); D. C. Johnston, Adv. Phys. \textbf{59},
803 (2010); J. Paglione and R. L. Greene, Nat. Phys. \textbf{6}, 645
(2010); P. C. Canfield and S. L. Bud'ko, Annu. Rev. Condens. Matter
Phys. \textbf{1}, 27 (2010); H.H. Wen and S. Li, Annu. Rev. Condens.
Matter Phys. \textbf{2}, 121 (2011); P. J. Hirschfeld, M. M. Korshunov,
and I. I. Mazin, Rep. Prog. Phys. \textbf{74}, 124508 (2011); A.V.
Chubukov, Annu. Rev. Condens. Matter Phys. \textbf{3}, 57 (2012).

\bibitem{Chubukov03} A. Abanov, A. V. Chubukov, and J. Schmalian,
Adv. Phys. \textbf{52}, 119 (2003).

\bibitem{Metlitski10} M. A. Metlitski and S. Sachdev, Phys. Rev.
B \textbf{82}, 075128 (2010).

\bibitem{Chubukov13} Y. Wang and A. V. Chubukov, Phys. Rev. Lett.
\textbf{110}, 127001 (2013)

\bibitem{Efetov12} H. Meier, C. Pepin and K. B. Efetov, arXiv:1210.3276

\bibitem{Zaanen11} J.-H. She, B. J. Overbosch, Y.-W. Sun, Y. Liu,
K. Schalm, J. A. Mydosh, and J. Zaanen, Phys. Rev. B \textbf{84},
144527 (2011).

\bibitem{Scalapino12} D. J. Scalapino, Rev. Mod. Phys. \textbf{84},
1383 (2012)

\bibitem{Vorontsov09} A. B. Vorontsov, M. G. Vavilov, and A. V. Chubukov,
Phys. Rev. B \textbf{79}, 060508(R) (2009).

\bibitem{Kasahara10} S. Kasahara et al., Phys. Rev. B \textbf{81},
184519 (2010).

\bibitem{FernandesPRB10} R. M. Fernandes, D. K. Pratt, W. Tian, J.
Zarestky, A. Kreyssig, S. Nandi, M. G. Kim, A. Thaler, N. Ni, P. C.
Canfield, R. J. McQueeney, J. Schmalian, and A. I. Goldman, Phys.
Rev. B \textbf{81}, 140501(R) (2010).

\bibitem{Vorontsov10} A. B. Vorontsov, M. G. Vavilov, and A. V. Chubukov,
Phys. Rev. B \textbf{81}, 174538 (2010).

\bibitem{Fernandes_Schmalian} R. M. Fernandes and J. Schmalian, Phys.
Rev. B \textbf{82}, 014521 (2010).

\bibitem{Hashimoto12} K. Hashimoto, K. Cho, T. Shibauchi, S. Kasahara,
Y. Mizukami, R. Katsumata, Y. Tsuruhara, T. Terashima, H. Ikeda, M.
A. Tanatar, H. Kitano, N. Salovich, R. W. Giannetta, P. Walmsley,
A. Carrington, R. Prozorov, and Y. Matsuda, Science \textbf{336},
1554 (2012); K. Hashimoto, Y. Mizukami, R. Katsumata, H. Shishido,
M. Yamashita, H. Ikeda, Y. Matsuda, J. A. Schlueter, J. D. Fletcher,
A. Carrington, D. Gnida, D. Kaczorowski, and T. Shibauchi Proc. Natl.
Acad. Sci. USA 110, 3293-3297 (2013).

\bibitem{Levchenko13} A. Levchenko, M. G. Vavilov, M. Khodas, and
A. V. Chubukov, arXiv:1212.5719; T. Nomoto and H. Ikeda, preprint.

\bibitem{Sachdev13} D. Chowdhury, B. Swingle, E. Berg, and S. Sachdev,
arXiv:1305.2918

\bibitem{Nandi10} S. Nandi, M. G. Kim, A. Kreyssig, R. M. Fernandes,
D. K. Pratt, A. Thaler, N. Ni, S. L. Bud'ko, P. C. Canfield, J. Schmalian,
R. J. McQueeney, and A. I. Goldman, Phys. Rev. Lett. \textbf{104},
057006 (2010).

\bibitem{Bohmer12} A. E. Bohmer, P. Burger, F. Hardy, T. Wolf, P.
Schweiss, R. Fromknecht, H. v. Lohneysen, C. Meingast, H. K. Mak,
R. Lortz, S. Kasahara, T. Terashima, T. Shibauchi, and Y. Matsuda,
Phys. Rev. B \textbf{86}, 094521 (2012).

\bibitem{Fernandes12} R. M. Fernandes, A. V. Chubukov, J. Knolle,
I. Eremin, and J. Schmalian, Phys. Rev. B \textbf{85}, 024534 (2012).

\bibitem{Fisher_RPP} I. R. Fisher, L. Degiorgi, and Z. X. Shen, Rep.
Prog. Phys. \textbf{74}, 124506 (2011).

\bibitem{Eremin10} I. Eremin and A. V. Chubukov, Phys. Rev. B \textbf{81},
024511 (2010).

\bibitem{Maiti10} S. Maiti and A. V. Chubukov, Phys. Rev. B \textbf{82},
214515 (2010).

\bibitem{FernandesPRL10} R. M. Fernandes, L. H. VanBebber, S. Bhattacharya,
P. Chandra, V. Keppens, D. Mandrus, M. A. McGuire, B. C. Sales, A.
S. Sefat, and J. Schmalian, Phys. Rev. Lett. \textbf{105}, 157003
(2010).

\bibitem{Kivelson} C. Fang, H. Yao, W.-F. Tsai, J. Hu, and S. A.
Kivelson, Phys. Rev. B \textbf{77}, 224509 (2008).

\bibitem{Sachdev} C. Xu, M. Muller, and S. Sachdev, Phys. Rev. B
\textbf{78}, 020501(R) (2008).

\bibitem{comm_a} Note that in our 3-band model SDW and SC orders
coexist at arbitrary $\delta_{0}/\delta_{2}$ when $\delta_{0}$ and
$\delta_{2}$ are small enough. In a two-band model, the coexistence
in this limit is possible only in some range of $\delta_{0}/\delta_{2}$~\cite{Vorontsov09,FernandesPRB10,Vorontsov10,Fernandes_Schmalian}.

\bibitem{Gorkov09} V. Barzykin and L.P. Gor'kov, Phys. Rev. B \textbf{79},
134510 (2009).

\bibitem{Qi09} Y. Qi and C. Xu, Phys. Rev. B \textbf{80}, 094402
(2009).

\bibitem{Cano10} A. Cano, M. Civelli, I. Eremin, and I. Paul, Phys.
Rev. B \textbf{82}, 020408(R) (2010).

\bibitem{batista} Y. Kamiya, N. Kawashima, and C. D. Batista, Phys.
Rev. B \textbf{84}, 214429 (2011).

\bibitem{Applegate11} R. Applegate, R. R. P. Singh, C.-C. Chen, and
T. P. Devereaux, Phys. Rev. B \textbf{85}, 054411 (2012).

\bibitem{lorenzana} G. Giovannetti, C. Ortix, M. Marsman, M. Capone,
J. van den Brink, and J. Lorenzana, Nature Comm. \textbf{2}, 398 (2011);
M. Capati, M. Grilli, and J. Lorenzana, Phys. Rev. B \textbf{84},
214520 (2011).

\bibitem{millis10} A. J. Millis, Phys. Rev. B \textbf{81}, 035117
(2010).

\bibitem{comm_aa} For $d_{eff}=4$, the magnitude of the jump in
$\varphi$ is generally of order of the upper cutoff $\Lambda$, and
gets smaller only for very large $\alpha$.

\bibitem{NMR_coexistence_Co1} Y. Laplace, J. Bobroff, F. Rullier-Albenque,
D. Colson, and A. Forget, Phys. Rev. B \textbf{80}, 140501(R) (2009).

\bibitem{NMR_coexistence_Co2} M.-H. Julien, H. Mayaffre, M. Horvatic,
C. Berthier, X. D. Zhang, W. Wu, G. F. Chen, N. L. Wang, and J. L.
Luo, EPL \textbf{87}, 37001 (2009).

\bibitem{NMR_coexistence_P} T. Iye, Y. Nakai, S. Kitagawa, K. Ishida,
S. Kasahara, T. Shibauchi, Y. Matsuda, and T. Terashima, J. Phys.
Soc. Jpn. \textbf{81}, 033701 (2012).

\bibitem{Gordon10} R. T. Gordon, H. Kim, N. Salovich, R. W. Giannetta,
R. M. Fernandes, V. G. Kogan, T. Prozorov, S. L. Bud'ko, P. C. Canfield,
M. A. Tanatar, and R. Prozorov, Phys. Rev. B 82, 054507 (2010).

\bibitem{Fernandes_Schmalian2} R. M. Fernandes and J. Schmalian,
Phys. Rev. B \textbf{82}, 014520 (2010).

\bibitem{Vavilov_penetration_depth} D. Kuzmanovski and M. G. Vavilov,
Supercond. Sci. Technol. 25, 084001 (2012).\end{thebibliography}

\begin{thebibliography}{10}
\bibitem{S_Maiti10} S. Maiti and A. V. Chubukov, Phys. Rev. B \textbf{82},
214515 (2010).

\bibitem{RMF} R. M. Fernandes, A. V. Chubukov, J. Knolle, I. Eremin,
J. Schmalian, Phys. Rev. B \textbf{85}, 024534 (2012).

\bibitem{2Delta} S. Maiti and A.V. Chubukov, Phys. Rev. B \textbf{83},
220508 (2011).

\bibitem{S_Vorontsov10} A. B. Vorontsov, M. G. Vavilov, and A. V. Chubukov,
Phys. Rev. B \textbf{81}, 174538 (2010).

\bibitem{subir} S. Sachdev, \emph{Quantum phase transitions}, Cambridge
University Press, 2011.

\bibitem{ref1} J. S. Langer, Ann. of Phys. \textbf{41}, 108 (1967). 

\bibitem{ref2} Hagen Kleinert, \emph{Path Integrals in Quantum Mechanics,
Statistics, Polymer Physics, and Financial Markets}, World Scientific
Publishing Company, 2006.\end{thebibliography}
\end{document}